\documentclass[12pt]{iopart}
\usepackage{graphicx}
\pdfoutput=1
\usepackage[dvipsnames]{xcolor}
\usepackage{caption}
\usepackage{todonotes}
\usepackage{subcaption}
\usepackage[squaren]{SIunits}
\usepackage{booktabs}
\usepackage{mhchem}
\usepackage{bbold}
\usepackage[normalem]{ulem}
\usepackage{hyperref}
\expandafter\let\csname equation*\endcsname\relax
\expandafter\let\csname endequation*\endcsname\relax
\usepackage{amsmath}
\let\vec\boldsymbol

\begin{document}

\title[]{Laser polarization control of ionization-injected electron beams and x-ray radiation in laser wakefield accelerators}

\author{Arghya Mukherjee$^\dag$ and Daniel Seipt}
\address{Helmholtz Institute Jena, Fröbelstieg 3, 07743, Jena, Germany}%
\address{GSI Helmholtzzentrum für Schwerionenforschung GmbH, Planckstra\ss{}e 1, 64291 Darmstadt, Germany}
\ead{a.mukherjee@hi-jena.gsi.de$^\dag$}

\vspace{10pt}
\date{\today}

\begin{abstract} 
In this paper we have studied the influence of the laser polarization on the dynamics of the ionization-injected electron beams and subsequently the properties of the emitted betatron radiation in laser wakefield accelerators (LWFAs). While ionizing by a strong field laser radiation, generated photo-electrons carry a residual transverse momentum in excess of the ionization potential via the above threshold ionization process. This above threshold ionization (ATI) momentum explicitly depends on the polarization state of the ionizing laser and eventually governs the dynamics of the electron beam trapped inside the wake potential. In order to systematically investigate the effect of the laser polarization, here, we have employed complete three dimensional (3-D) Particle-in-Cell (PIC) simulations in the nonlinear bubble regime of the LWFAs. We focus, in particular, on the effects the laser polarization has on the ionization injection mechanism, and how these features affect the final beam properties, such as, beam charge, energy, energy spread and transverse emittance. We have also found that as the laser polarization gradually changes from linear to circular, the helicity of the electron trajectory, and hence the angular momentum carried by the beam increases significantly. Studies have been further extended to reveal the effect of the laser polarization on the radiation emitted by the accelerated electrons. The far field radiation spectra have been calculated for the linear (LP) and circular polarization (CP) states of the laser. It has been shown that the spatial distributions and the polarization properties (Stokes parameters) of the emitted radiation for the above two cases are substantially different. Therefore, our study provides a facile and efficient alternative to regulate the properties of the accelerated electron beams and x-ray radiation in LWFAs, utilizing ionization injection mechanism.
\end{abstract}

%
%
%
%
%

\newpage

\section{Introduction}  \label{sect:intro}

    The physics of laser wakefield acceleration (LWFA) of electrons in a plasma has received extensive attention in last few decades \cite{tajima, esarey, esarey_rmp, Tajima2020, joshi, Albert_2021, fa} since their first theoretical proposal in 1979 \cite{tajima}.  In LWFA, a relativistically intense laser pulse is focused into an undercritical plasma where it is exciting a plasma wake (`bubble') with strong longitudinal electrical field of up to $\unit{100}{\giga\volt\per\metre}$ \cite{esarey_rmp}. By trapping electrons in the wake field the longitudinal fields are used to accelerate electrons to GeV on centimetre scales, which allows for very compact accelerators compared to conventional RF technology.
    LWFA experiments have already shown their capability to produce monoenergetic (multi-)GeV electron beams with an energy spread only of a few percent \cite{malka, Mangles2004, Geddes2004, Faure2004, Leemans2006, l14, kim13, gon19, ke21, foester22}.
    
    Current LWFA research focuses mostly on improving the electron beam quality and operating LWFA's at higher repetition rate. One key aspect is in better reproducibility and control of important properties such as beam charge, energy and pointing stability \cite{maier_decoding_2020, seidel2022polarization, jalas_tuning_2023}, or improving the beam quality (e.g. the beam emittance or energy spread) \cite{reg2, jalas_bayesian_2021, kirchen_optimal_2021}; often by employing machine learning techniques and automation \cite{dopp_data-driven_2023, shalloo_automation_2020, dann_laser_2019}. These improvements are required in view, e.g., of the necessary staging of plasma accelerators \cite{steinke_staging, kim13, foester22, reg2} to reach electron energies in excess of a TeV and sufficient luminosity for high-energy physics collider experiments \cite{benedetti2022whitepaper,roser_feasibility_2023}.    Precise control of the electron properties is naturally difficult due to the transient nature of the plasma in which the LWFA process takes place. This puts demanding requirements on the laser and plasma target to achieve the desired beam parameters, especially for controlling the beam's phase-space properties. The primary opportunity to control the beam properties can be found in the electron injection process.

    At high enough laser intensity, the plasma wave breaks and electrons are injected into the bubble \cite{esarey_rmp,kalmykov_injection_2006,schroeder_trapping_2006,Benedetti:POP2013,kuschel_controlling_2018}. Recently it has also been shown that the self-injection threshold can be lowered by using circularly polarized laser pulses \cite{Ma2020,ma_effects_2021}. However, using self-injection it is very hard to control the electron beam properties as one desires, because of the electron injection typically changes from shot-to-shot in this scheme \cite{Dpp2017}. For this reason, various techniques have been developed for a controlled electron injection into plasma wakefields: density tailoring and down-ramp-injection schemes \cite{dt1, dt2, dt3, ekerfelt_tunable_2017,ullmann_all-optical_2021}, colliding pulse-injection \cite{mp1,faure2006colliding}, external magnetic fields schemes \cite{mf1, mf2}, and ionization injection \cite{ii0, Pak2010_ionization, ii2,jalas_bayesian_2021} to name a few. The aim of this paper is to investigate, specifically for the case of ionization injection, how the injected electron beam properties can be controlled via the laser polarization.

    For ionization injection, a small percentage of high-$Z$ injector gas (e.g.~Nitrogen, Argon, ...) is mixed to the low-$Z$ Hydrogen or Helium target gas \cite{Pak2010_ionization}. Due to the higher ionization potential (IP), the inner shell electrons of the high-Z injector gas get tunnel-ionized near the peak of the laser pulse inside the first plasma bubble. The ionized electrons then slip backward relative to the wake and get trapped inside the wake potential. At this point, the initial conditions for the acceleration are determined by the outcome of the tunnel-ionization process which we can control via the laser polarization. The acceleration of electron bunches of a few fs duration with narrow energy spreads and low emittances have been demonstrated by employing ionization injection scheme in a variety of LWFA regimes \cite{ii2, reg2, reg3, reg4, reg5, reg6, reg7, reg8, reg9, reg10, reg11}.
    
    While being accelerated, the bunch electrons perform transverse betatron oscillations due to the focusing force produced by the immobile ions inside the cavity \cite{igor}. The emitted betatron radiation from the accelerating electrons is typically in the x-ray range \cite{rad1, rad2, rad3}, with its key properties (such as brightness, spectral features, spatial distribution, polarization etc.) fully determined by the electron energy and the transverse phase-space properties of the trapped electrons \cite{rad4, popprad, Schnell2013, rmprad, Dpp2017}. Betatron radiation has found a number of applications, e.g. in the study of ultra-fast physical, chemical and biological processes owing to their capability of resolving the structure and dynamics of matters on the atomic and molecular scales \cite{app1, app2, app3, app4, app5, app6, app7}. A precise steering of the initially injected electrons, and hence their betatron orbits, also serves to control the properties of the emitted betatron radiation in LWFA \cite{Schnell2013,Dpp2017}.

    In this paper we study how the accelerated electron beam and betatron radiation properties can be controlled via the laser polarization in ionization injection schemes of LWFA. The main idea is to control the initial conditions of the electrons for the subsequent acceleration.  During the ionization process, generated photo-electrons carry a residual transverse momentum in excess of the ionization potential via the above threshold ionization (ATI) process. The residual ATI momentum has a strong dependence on the polarization state of the ionizing laser, owing to the conservation of the transverse momentum during electron's interaction with the laser. Here we primarily focus on the effects of ATI momentum, and hence the laser polarization, on the electron beam trajectories and the final beam properties, such as, beam charge, energy, energy spread and transverse emittance. By performing three dimensional (3-D) Particle-in-Cell (PIC) simulations we demonstrate that an initial change in the laser polarization, and hence in the ATI momentum, dramatically affects the dynamics of the trapped electron orbits and hence the final accelerated beam characteristics and the emitted betatron x-ray radiation.

    This paper is organized as follows: In section~\ref{sect:ionization}, we first review the strong-field ionization process in a strong laser field with arbitrary polarization and identify the effective regimes for the photo-electron spectra. Then in Section~\ref{sect:pic} we present PIC simulation results to explore the effect of the laser polarization on the injected electron trajectories inside the bubble and on the electron beam properties. Section~\ref{sect:radiation} is devoted to calculating the betatron radiation emitted during the acceleration process. Here we present a comparative study on the radiated energy spectra and spatial distribution, and discuss the polarization properties via Stokes parameters. Finally, in Section \ref{sect:conclusion}, we summarize our work.

\section{Theory of strong-field ionization in a laser pulse with arbitrary polarization}   \label{sect:ionization}

    The aim of our present work is to demonstrate that the  characteristics of the ionization injected electron beams and their radiation properties can be controlled by the polarization state of the ionizing laser. Here the outcome of the ionization process by the laser radiation determines the initial conditions for the electrons trapped inside the bubble and hence the control of ionization via polarization allows one to control the betatron orbits of the electrons during the acceleration process. Therefore, in this section we briefly discuss the tunneling ionization process of atomic electrons by a strong  laser pulse with arbitrary polarization.

    The phenomenology of the strong field ionization process via the ATI mechanism in ionization injection can be understood as a 2-step process \cite{Corkum1989}. In the first step the electron tunnel ionizes instantaneously from a deeply bound state to the continuum, where it is assumed to have zero velocity initially. In the second step, in the continuum, the electron's transverse canonical momentum is conserved, $\vec p_\perp - e \vec A = const$. Thus, after the electron has left the laser pulse the electron has gained the residual transverse momentum $\vec p_\perp =m_ec \vec a_i$, where $\vec a_i = e \vec A_i/m_ec$ is the normalized laser vector potential at ionization.

    As an immediate consequence of this the electron momentum distributions depend on the polarization of the laser pulse.
    For instance, for a linearly polarized laser pulse, strong-field ionization mostly occurs at the crests of the oscillating electric field, which leads very small momentum because $\vec a_i \approx 0$; a final momentum is only possible for a phase mismatch $\varphi_0$ \cite{schroeder_thermal_2014} which, however, is suppressed due to the exponential dependence of the tunnel ionization rates on the electric field strength \cite{Corkum1989,sp4,sp_book}.
    Contrary, for circular polarization, the magnitude of the electric field does not oscillate, instead it follows the pulse envelope. The vector potential is $90$ degrees out of phase with the electric field, but its magnitude at the moment of ionization, $\vec a_i$, is non-vanishing yielding large transverse electron momentum.  Moreover, for LP the electrons are ionized in ultra-short bursts, while for CP the ionization is more continuous \cite{ma_effects_2021}. Therefore, depending upon the polarization state of the laser and the IP of the atomic electrons, the ATI momenta exhibit a variety of distributions, which ultimately affects the accelerated electron beam properties.

    To predict the size of the effect we can estimate the vector potential at ionization $a_i$ using the Barrier suppression field,~$E_{BSI}=I_P^2/(4Ze\alpha\hbar c)$, where $I_P$ is the ionization potential of the electron, $Z$ is the charge number of the final ion and $\alpha$ is the fine structure constant. For instance, the 1s electron of Helium has an $I_P=\unit{54.4}{\electronvolt}$, yielding thus $a_i=0.064$ for a circular polarized laser. Even though this value is quite small, it was observed that this CP effect can reduce the self-injection threshold in LWFA \cite{Ma2020,ma_effects_2021}. For higher-$Z$ atoms, the corresponding values of $a_i$ are much larger. For the 1s electron of Carbon we have $I_P=\unit{490}{\electronvolt}$ and $a_i=1.7$, which corresponds to an ATI momentum of $p_\perp c = 879$ keV. For Nitrogen, the corresponding values are $I_P=\unit{667}{\electronvolt}$, $a_i=2.7$ and $p_\perp c = 1397$ keV.

    The detailed energy and momentum distributions of the photo-electrons ionized in a light wave of various polarization have been studied in the literature \cite{sp1, sp2, sp3, sp4, sp5, sp6, sp_book}. Here we only summarize the main features of the photo-electrons transverse momentum distribution subject to the polarization state of the ionizing light wave \cite{Mur2001, pop2004}. Let's consider a plane wave laser propagating along the $x$-direction, with its polarization parameterized by the ellipticity parameter $\epsilon$, and having the electric field of the following form, 
    \begin{equation}
    \vec{E}_L = \frac{E_{0}\cos (\omega_Lt - k_Lx)}{(1 + \epsilon ^2)^{1/2}}\hat{y} +  \frac{\epsilon E_{0}\sin (\omega_Lt - k_Lx)}{(1 + \epsilon ^2)^{1/2}}\hat{z} \label{eq0} 
    \end{equation} 
    where $\epsilon=0$ represents linear polarization (LP) and $\epsilon=1$ corresponds to circular polarization (CP). $\omega _L$ and $k_L$ respectively be the frequency and the wave number of the laser. Here $\vec E_L$ has been defined to have the same intensity for a given $E_0$ irrespective of the laser polarization, but the peak field explicitly depends on $\epsilon$ and can be written as $\widehat{|\vec E_L|} = E_0/\sqrt{1+\epsilon^2}$. Similarly, the peak value of the normalized vector potential of the laser can be defined as $\hat a_0 = a_0/\sqrt{1+\epsilon^2}$, where $a_0  = eA_0/m_ec = eE_0/m_e\omega _Lc$.

    Now, the course of the strong field ionization process can be characterized by the value of the Keldysh parameter $\gamma _K = \omega_L\sqrt{2m_eI_P}/e|\vec {E}_L|$
    \cite{kel}, which essentially distinguishes two different regimes. For $\gamma_K \gg 1$ one has multi-photon ionization and for stronger fields, $\gamma_K \ll 1$, ionization occurs in the quasi-static strong-field (tunneling) regime. The latter one is relevant for LWFA ionization injection. For even stronger fields the electrons don't have to tunnel, they may classically escape over the barrier.
    
    The effect of the polarization on the photo-electrons' transverse momentum distribution $\vec p_\perp = (p_y, p_z)$ is characterized by an another dimensionless parameter, the so-called reduced electric field $\chi = 1/2K_0\gamma_K = e\hbar m_e |\vec {E}_L| /(2m_eI_P)^{3/2}$ 
    , where $K_0=I_P/\hbar \omega_L$ is the minimum number of laser photons required for the process \cite{Mur2001}.

    Depending on the values of $\epsilon$ and $\chi$ we can distinguish the following three categories of transverse momentum distributions in the three different regimes (assuming the longitudinal momentum of the ionized electrons $p_x = 0$ during the ionization) \cite{Mur2001, pop2004}:

\begin{enumerate}
    \item LP-like regime --- for $\epsilon < \sqrt{\chi}$ --- photo-electrons transverse momentum distribution will be a double peaked Gaussian with peaks at $p_y = 0, p_z \neq 0$ and distributed along the major axis of the polarization ellipse (the $p_y$-axis in our case). The normalized distance between the two Gaussian peaks along the $p_z$-axis is equal to the twice of the $z$ component of the normalized vector potential of the laser. Therefore, for $\epsilon \approx 0$ the double peaked Gaussian distributions degenerate to a single peaked Gaussian with a maximum at $p_y, p_z = 0$. 
    \item EP-like regime --- for $\sqrt{\chi} < \epsilon < 1 - \chi$ --- we essentially get an elliptical distribution of the transverse momentum of the generated photo-electrons. The normalized major and minor radius of the ellipse correspond to the respective values of the $y$ and $z$ components of the normalized vector potential's peak amplitude.
    \item CP-like regime --- for $\epsilon > 1 - \chi$, the photo-electrons transverse momentum distribution will appear as a circular ring on the transverse plane with a normalized radius equal to the normalized vector potential's peak amplitude.
\end{enumerate}

\begin{figure}
    \centering
    \includegraphics[width=0.69\textwidth]{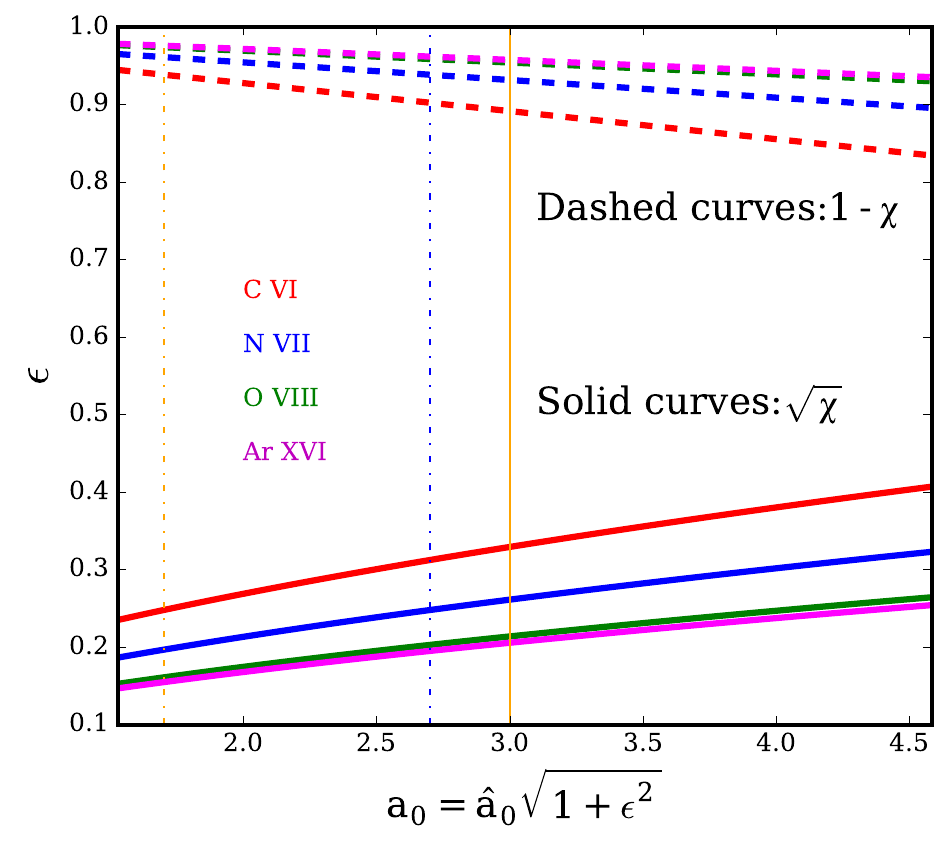}
    \caption{Variations of $\sqrt{\chi}$ and $1- \chi$ as a function of normalized vector potential of the ionizing laser given by Eqn.~\eqref{eq0}. The vertical dashed yellow and blue line respectively represents the minimum vector potential values ($a_i$) corresponding to the barrier suppression field for $\ce{C}$ VI and $\ce{N}$ VII, respectively.}
    \label{species}
\end{figure}
Now, $\sqrt{\chi}$ and $1- \chi$, both depend on the $I_P$ of the atomic electrons in the injector gas species and $a_0$. The variations of $\sqrt{\chi}$ and $1- \chi$ as a function of $a_0$ have been depicted in Fig.~\ref{species} for different injector gas species. The solid and dashed lines respectively represent the quantities $\sqrt{\chi}$ and $1- \chi$ for Hydrogen-like Carbon (red), Nitrogen (blue), Oxygen (green) and Argon (magenta). For a fixed value of laser intensity ($a_0$) and a fixed injector gas species, if we change the ellipticity parameter $\epsilon$, for $\epsilon$ values below the solid curve the photo-electron transverse momentum distribution will be double peaked (LP-like regime). In the regime between the solid line and dashed line (of the same colour/species) the distribution will be an elongated ellipse (EP-like regime) and finally for the values of $\epsilon$ above the dashed line, the distribution is almost like a circular ring (CP-like regime). The vertical dashed yellow and blue line respectively represents the minimum vector potential values $a_i$  required for barrier suppression ionization of Hydrogen-like Carbon and Nitrogen. The corresponding values of $a_i$ (determined via the barrier suppression ionization model) for the various ions (using spectroscopic notation) $\ce{C}$ V, $\ce{C}$ VI, $\ce{N}$ VI, $\ce{N}$ VII, $\ce{O}$ VII and $\ce{O}$ VIII are given by 1.3, 1.7, 2.2, 2.7, 3.4 and 4.1, respectively. By comparing the $a_i$ values for different injector gas species we can notice that the lowest intensity be required for Carbon. 

Here, we would like to emphasize that although the ionization dynamics is governed by the peak field (or, the peak vector potential $\hat a_0$), plasma dynamics is essentially governed by the time-averaged vector potential, which is polarization independent and in the normalized form can be written as $\bar a_0 = a_0/\sqrt{2} $. In addition to this, the efficiency of ionization injection scheme crucially depends on the matching between the laser peak intensity and the IP level of the injector, so that most of the inner shell electrons get ionized near the peak of the laser. By our choice of injector gas and $a_0$ we have made sure that for all the polarization the lowest lying ionization level can be ionized. In other words, if the lowest $I_P$ is too large and $a_0$ is relatively small, then for increasing $\epsilon$ the lowest lying electrons cannot be accessed which would have tremendous consequences on the injected charge \cite{Hafezi}. An evidence of this can be seen in a very recent publication \cite{Hafezi} where it has been found that for a peak laser amplitude $a_0 = 1.4$, $\ce{N}$ VI electrons are formed at the peak of the laser for LP case, whereas almost no $\ce{N}$ VI electrons are generated for CP case. Here, by our choice of the injector gas and $a_0$, we mitigate this effect, thus focusing on the effect
of the ATI momentum (laser polarization). Therefore, if we take a laser with a vector potential $a_0 =  3$ (shown by the vertical solid yellow line in Fig.~\ref{species}) and choose Carbon as injector, we can ionize the K-shell electrons for all polarization and it will additionally provide us to access all the three different regimes (LP/EP/CP-like) of photo-electron's momentum distributions, which can be obtained by gradually tuning the laser polarization, without changing its time-averaged intensity. In the next section we show that the variation of the ionized photo-electrons transverse momentum distribution with the laser polarization $\epsilon$, presented in the above summary, is directly linked to the variation of the transverse emittance of the final accelerated electron beams with respect to the same parameter $\epsilon$.

\section{Particle-in-Cell simulation results}    \label{sect:pic}

    \subsection{Simulation set-up details}

    In this paper, complete 3D PIC simulations have been carried out in cartesian geometry with the open-source code SMILEI \cite{DEROUILLAT2018351}. We examine the effect of laser polarization on the ionization injected electron beams produced by a single, intense laser pulse propagating through a background plasma with with a mixture of Carbon as injector species. Experimentally this could be realized as a mixture of e.g.~Hydrogen and Methane gas. The background plasma is initialized in the simulation as pre-ionized plasma medium with a $\unit{100}{\micro\metre}$ long linear entry ramp and then with uniform density $n_p$. The injector species is localized to a $\unit{100}{\micro\metre}$ long region following the entry ramp, where the Carbon is initialized as neutral atoms, with the ADK model used for ionization \cite{DEROUILLAT2018351}. The Carbon number density is $0.005n_p$, 
    which is an optimal injector gas concentration seen in a previous experiment \cite{reg7}.

    To establish the effect of the laser polarization, we consider a Gaussian laser pulse of arbitrary ellipticity $\epsilon$ (defined in Eqn.~\eqref{eq0}), travelling along the $x$-direction, with the vacuum focus location at $x_f=\unit{180}{\micro\meter}$. The simulations have been carried out with a normalized laser amplitude $a_0 = 3 $. The laser has a longitudinal FWHM  = $\unit{10}{\micro\meter}$ and a transverse spot size $w_0 = \unit{15}{\micro\meter}$. Plasma density $n_p$ is taken at a particular value matched with the corresponding laser intensity and the transverse spot size \cite{lu}, calculated using $n_p = 4a_0/w_0^2 \approx 0.00086\times n_c$ in all the simulation runs. Here $n_c = \unit{1.745 \times 10^{21}}{\per\centi\meter^3}$ is the critical density for the $\lambda_L = \unit{0.8}{\micro\meter}$ laser, employed in our work.  The simulation box has dimensions $\ell_x \times \ell_y \times \ell_z = \unit{51.2 \times 76.8 \times 76.8}{\micro\meter^3}$ with a resolution of $\unit{0.04}{\micro\meter}$ and $\unit{0.4}{\micro\meter}$ along the longitudinal and transverse directions, respectively. We employ a moving window simulation.

    \subsection{Electron beam dynamics for LP and CP cases}
    
    In this section we present a comparative study of the electron beam dynamics between LP and CP cases. The beam properties have been measured at $t =$ $\unit{2}{\pico\second}$, which corresponds to a distance of $\sim$ \unit{0.6}{\milli\meter}. This distance is sufficient to compare the quality of the electron beam as the driver pulse has travelled adequate distance from the ionization region such that the injection process has finished. The beam charge and normalized transverse emittances are expected to be constant during the subsequent acceleration process.
    
\begin{figure}
     \centering
     \begin{subfigure}[b]{0.49\textwidth}
         \centering
         \includegraphics[width=\textwidth]{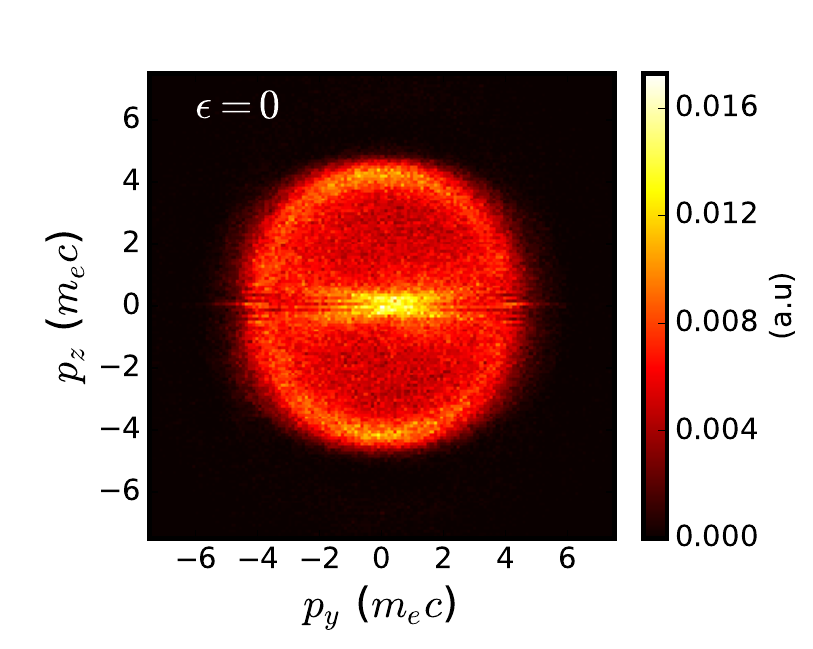}
         \caption{Linear polarization: along $y$}
         \label{pp1}
     \end{subfigure}
     \hfill
     \begin{subfigure}[b]{0.49\textwidth}
         \centering
         \includegraphics[width=\textwidth]{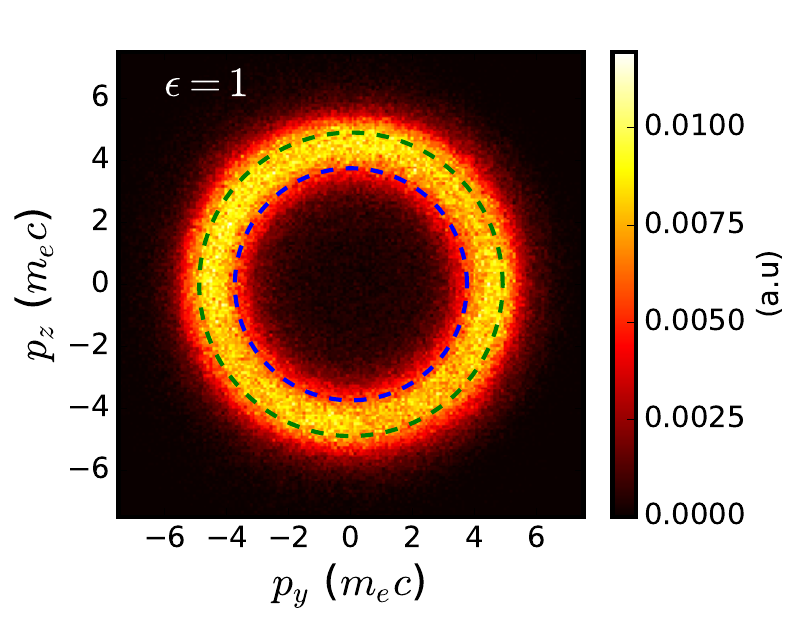}
         \caption{Circular polarization}
         \label{pp2}
     \end{subfigure}
     \caption{Transverse momentum distributions of the trapped electrons for two different laser polarization: left - LP($\epsilon = 0$, laser is polarized along $y$-direction), right - CP ($\epsilon = 1$). The solid circles represent the theoretical average radius calculated using Eq.~\eqref{eq2} for $\ce{C}$ V (blue) and $\ce{C}$ VI (green) electrons.}

     \label{figpp}
\end{figure}

Fig.~\ref{figpp} shows 2D colormaps of transverse momentum distributions of the trapped electrons for LP (left - Fig.~\ref{pp1}) and CP cases (right - Fig.~\ref{pp2}). Here we see that for LP case a large fraction of electrons is situated near the origin of the transverse momentum plane, with the distribution being elongated along the laser polarization axis.
Another fraction of electrons are distributed almost uniformly over a disk in the transverse plane. 
Contrary, for the CP case, the electrons' transverse momentum distribution exhibits a ring shaped structure. This difference is manifested by the difference in the photoelectrons momentum distributions at the instant of their ionization and their interaction with the laser during their passage through the bubble before getting trapped. It was mentioned in Section \ref{sect:ionization} that when electrons are ionized by a LP pulse, their transverse momentum exhibits a peak at zero transverse momenta values and the distribution is elongated along the major axis component ($p_y$-axis for momentum) of the laser polarization. For LP case a large fraction of electrons gets ionized on-axis at the crest of the electric field with a zero ATI momentum and few electrons generated via off-axis ionization have a nonzero ATI momentum, depending upon the phase of the laser electric field at the ionization point. As a result the ponderomotive drifts of the electrons during their passage through the cavity are not always directed along the direction of the laser electric field and during the acceleration process electrons transverse momentum distribution looks substantially different as described by the ionization model. But for CP case, transverse momentum distribution of the photoelectrons appears as a ring with a normalized radius $a_i$ and during their passage though the ion cavity these electrons receive a ponderomotive drift along the direction of the laser polarization. Therefore while performing betatron oscillations transverse momentum distribution manages to retain its ring shape structure with an increased radius at subsequent times of acceleration. 

The average radius of the ring at any arbitrary time during acceleration can be estimated theoretically. In our simulation a large fraction of $\ce{C}$ V and $\ce{C}$ VI electrons get ionized near the peak of the laser with a respective normalized transverse ATI momenta $1.3$ and $1.7$, inside the first plasma bubble (the outer states atomic electrons $\ce{C}$ I - $\ce{C}$ IV slip over the potential and are not trapped because of lower ionization threshold). It is well known that in the blowout regime the amplitude of the electron transverse momentum varies as $\gamma ^{1/4}$, where $\gamma$ represents the Lorentz factor \cite{Pukhov2002}. If $\gamma_i$ represents the Lorentz factor of the photo-electrons at the instant of ionization then the average radius of the ring in the transverse momentum space ($p_\perp(t)$) at any arbitrary time can be written as 
\begin{equation}
\frac{p_\perp(t)}{m_ec} = a_i \left(\frac{\bar \gamma(t)}{\gamma_i}\right)^{1/4}    \label{eq2}    
\end{equation}
where, $\bar \gamma(t)$ is the average Lorentz factor of the trapped electrons at the instant when the radius is being calculated and $\gamma _i$ has been calculated using $\gamma _i = \sqrt{1 + a_i^2}$. A comparison of our theoretical estimations have been presented by blue and green circles in Fig.~\ref{pp2}. The respective radius of the blue circle and the green circle represents the average radius calculated using the transverse ATI momentum of $\ce{C}$ V ($a_i = 1.3$) and $\ce{C}$ VI ($a_i = 1.7$) electrons.
We see that the average radius calculated using the ATI momentum of $\ce{C}$ V and $\ce{C}$ VI electrons are respectively close to the inner and outer radius of the ring and thus support our theoretical estimation.
\begin{figure}
     \centering
     \begin{subfigure}[b]{0.49\textwidth}
         \centering
         \includegraphics[width=\textwidth]{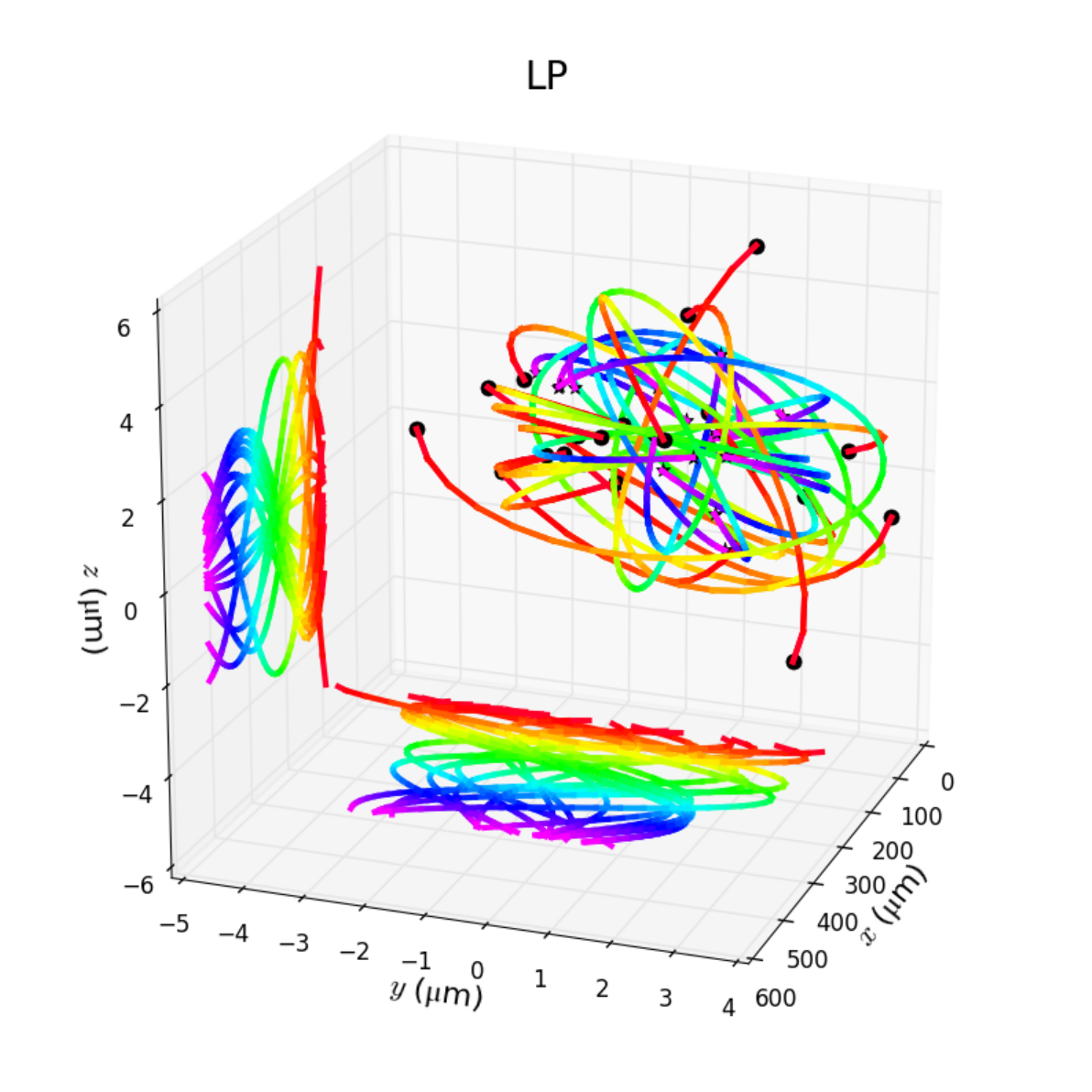}
         \caption{Linear polarization: along $y$}
         \label{3pp1}
     \end{subfigure}
     \hfill
     \begin{subfigure}[b]{0.49\textwidth}
         \centering
         \includegraphics[width=\textwidth]{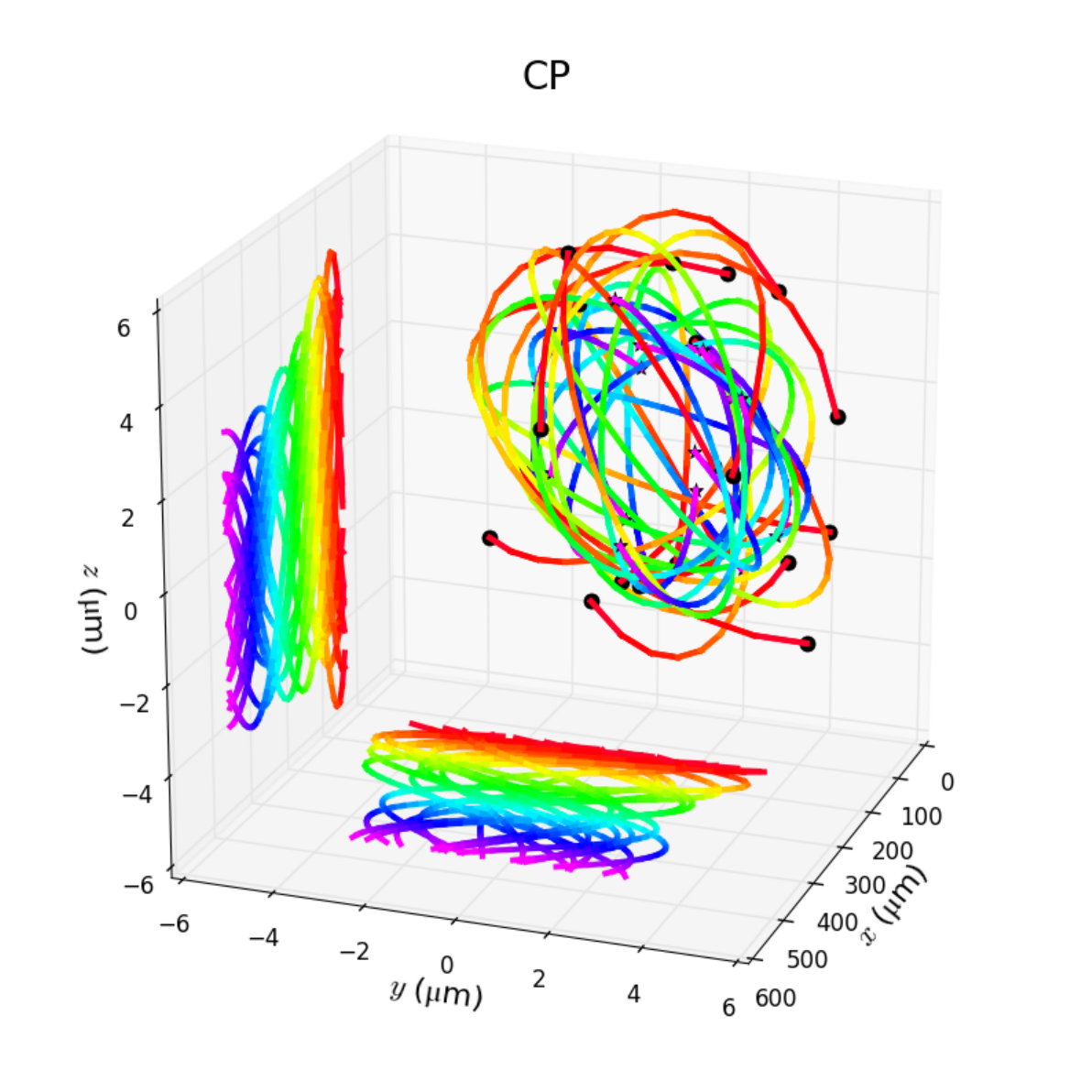}
         \caption{Circular polarization}
         \label{3pp2}
     \end{subfigure}
     \caption{
     Trajectories of few trapped electrons for LP (left) and CP cases (right). Here the color represents the time. Block dots are shown on the $y-z$ plane to demonstrate the initial starting points of the electrons. The color represents the time.}
        \label{fig3pp}
\end{figure}

The initial conditions of the trapped electrons for the acceleration process can be explicitly controlled by changing the polarization state of the laser, which ultimately changes the entire dynamics of the accelerated electron beam. In order to emphasize this fact now we show the trajectories of few trapped electrons, carrying maximum energy inside the beam, for the above two cases. These have been illustrated in Fig.~\ref{fig3pp}. For the sake of comparison we have displayed the trajectories along $x-y$, $x-z$ and $y-z$ planes. In both the cases electrons are performing betatron oscillations during the acceleration period which can be seen in the $x-y$ and $x-z$ planes. A closer inspection of the trajectories on the $x-z$ plane reveals that the betatron oscillation amplitude is higher for CP case as compared to the same for LP case. It occurs due to the presence of the laser electric field component along the $z$ direction for CP case, which ultimately leads to an enhancement of the normalized RMS emittance along $z$ direction. Now we compare the trajectories shown in the $y-z$ plane. The initial starting points of the betatron oscillations have been marked with black dots. In this Fig we see that for LP case, electrons initial transverse positions are randomly distributed over the $y-z$ plane before the betatron oscillation starts as the ponderomotive drift is not aligned to a specific direction. Consequently, different electrons have different ellipticities and orientation angle in their transverse trajectories depending upon their initial momenta and positions \cite{Dpp2017}. While for CP case we see that the initial positions (black dots) are distributed nearly over a periphery of a circle. In this case electrons have a well defined ATI momentum ($m_e c \vec a_i$) 
and ponderomotive drift is more aligned to the direction of laser electric field oscillation. 
\begin{figure}
     \centering
     \begin{subfigure}[b]{0.49\textwidth}
         \centering
         \includegraphics[width=\textwidth]{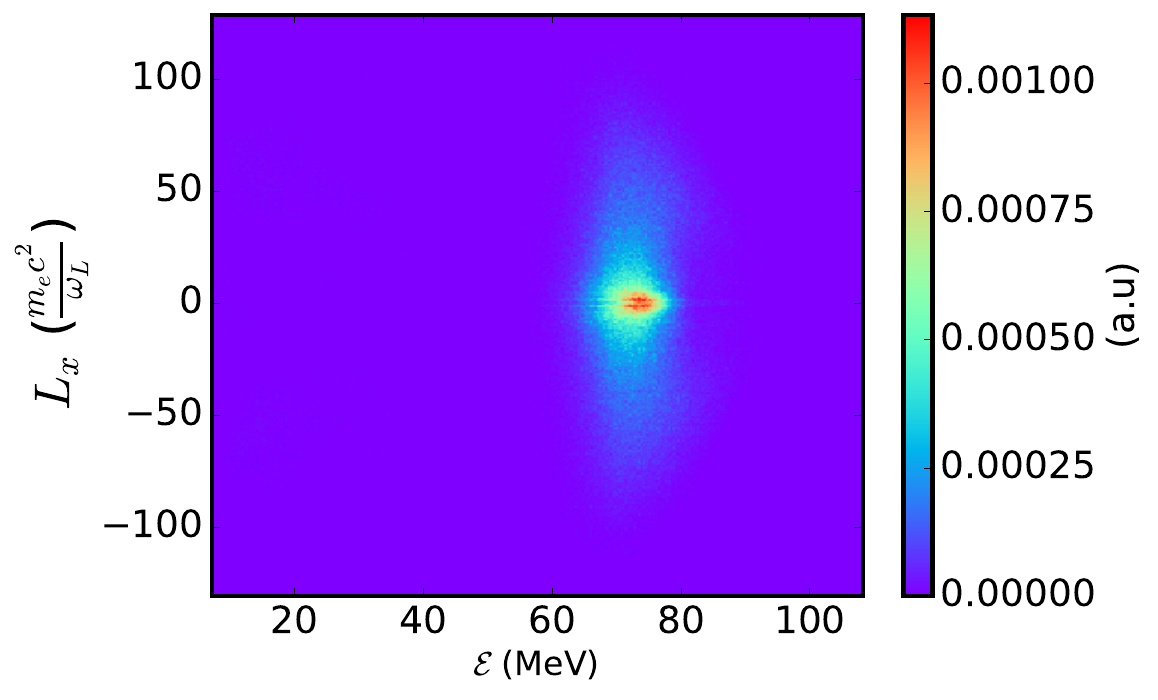}
         \caption{Linear polarization along $y$-axis; $\epsilon = 0$.}
         \label{6a}
     \end{subfigure}
     \hfill
     \begin{subfigure}[b]{0.49\textwidth}
         \centering
         \includegraphics[width=\textwidth]{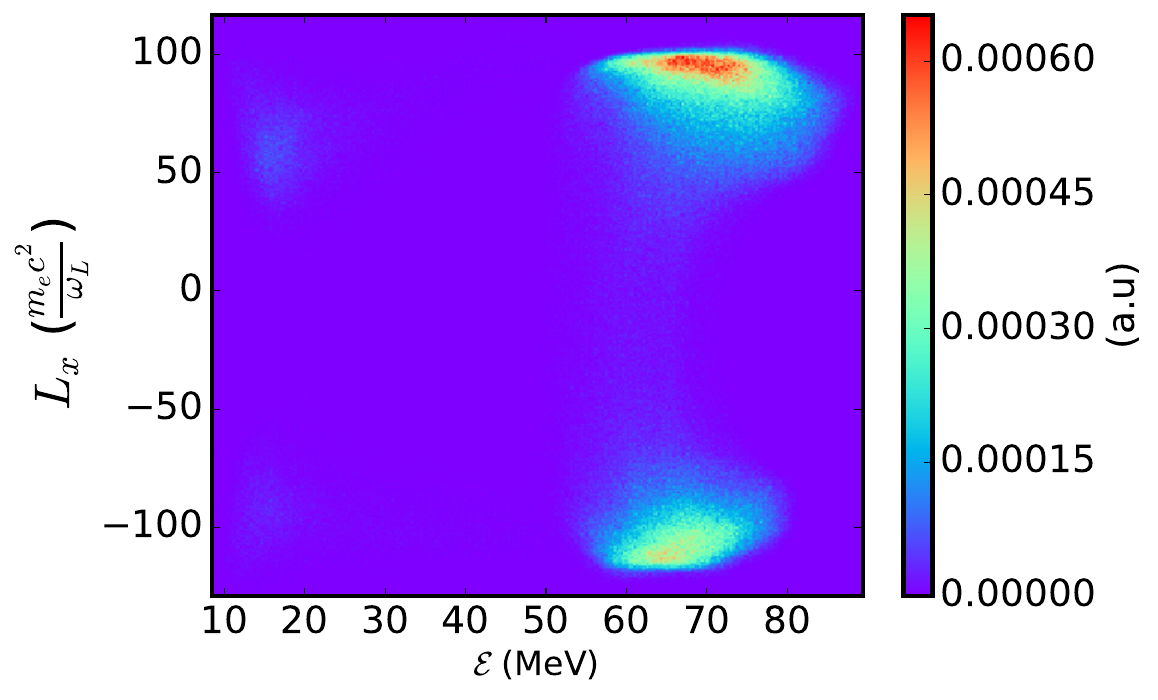}
         \caption{Circular polarization; $\epsilon = 1$.}
         \label{6b}
     \end{subfigure}
      \hfill
    \caption{2D colormap of energy ($\mathcal{E}$) and longitudinal angular momentum ($L_x$) of the electron beam  for LP (left) and CP (right) cases at $t=\unit{2}{\pico\second}$ after the start of the simulation.}
        \label{Energy2D}
\end{figure}

Another remarkable influence of the laser polarization on the accelerated electron beams can be seen in Fig.~\ref{Energy2D}. In this Fig. we have depicted 2D colormaps of the longitudinal angular momentum $L_x = yp_z - zp_y$ and the energy of the electron beam for LP (left) and CP (right) cases. The respective values of the peak energy (where the energy distribution has its peak) of the electron beam for the LP and CP cases are \unit{81}{\mega\electronvolt} and \unit{73}{\mega\electronvolt}, which is almost of the same order but the main difference between the two cases can be seen in the angular momentum distributions. From this Fig. we observe that for LP case, the electrons carrying the peak energy inside the beam have zero angular momentum. But for CP case the most energetic electrons inside the beam carry a nonzero angular momentum, distributed near $\omega _LL_x/m_ec^2 \approx \pm 100$. Physically, as the CP laser pulse carries a nonzero spin angular momentum (SAM) due to the rotation of the electric field, this SAM can be transferred efficiently to the electrons via the ATI momentum during the ionization process and subsequently via the ponderomotive drift during their passage thought the first bubble. The presence of this angular momentum is physically manifested by the helicity in the electron trajectories. Therefore our study also reflects that it is also possible to generated electron beams with an angular momentum by employing the ionization-injection method with a CP laser. Such electrons beams with a nonzero angular momentum are desired in condense matter spectroscopy, new electron microscopes and also in vortex radiation beam generations \cite{am1, am2, am3, am4, am5, am6}. In the next section we will see that, the tuning of the angular momentum of the ionization-injected electron beams can be performed by regulating the laser polarization. 
\subsection{Effects of laser polarization on the accelerated electron beams properties}
\begin{figure}
     \centering
     \begin{subfigure}[]{0.3\textwidth}
         \centering
         \includegraphics[width=\textwidth]{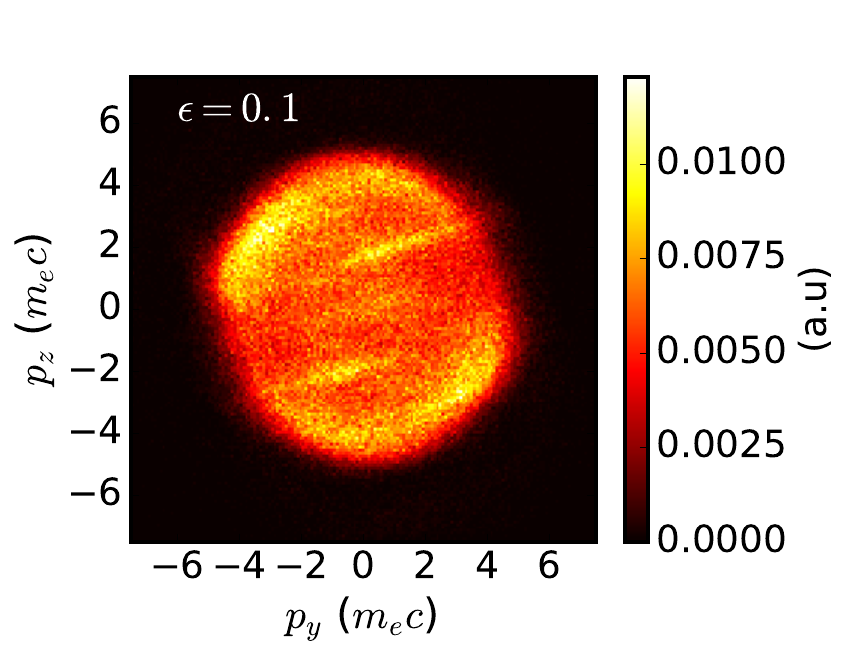}
         \label{ep1}
     \end{subfigure}
     \hfill
     \begin{subfigure}[]{0.3\textwidth}
         \centering
         \includegraphics[width=\textwidth]{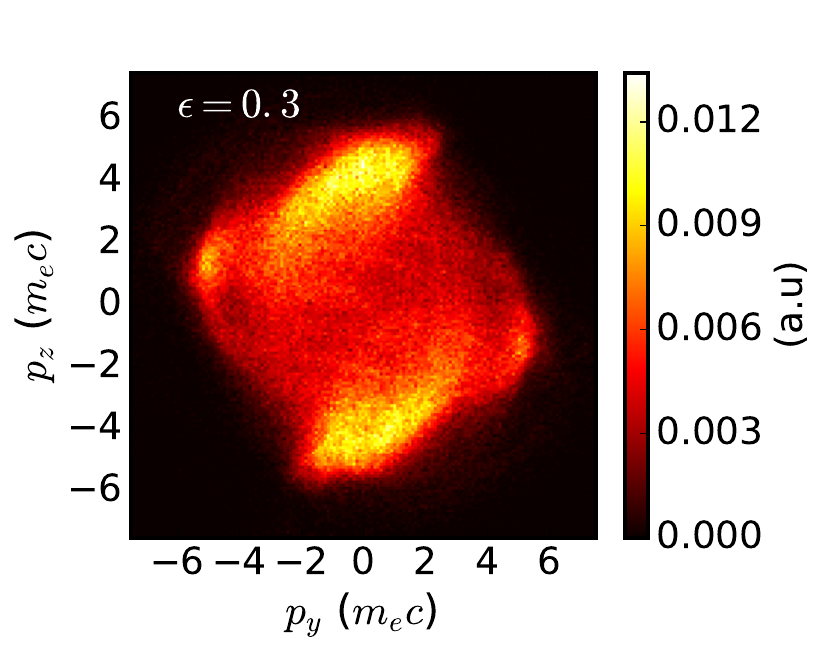}
         \label{ep2}
     \end{subfigure}
     \hfill
     \begin{subfigure}[]{0.3\textwidth}
         \centering
         \includegraphics[width=\textwidth]{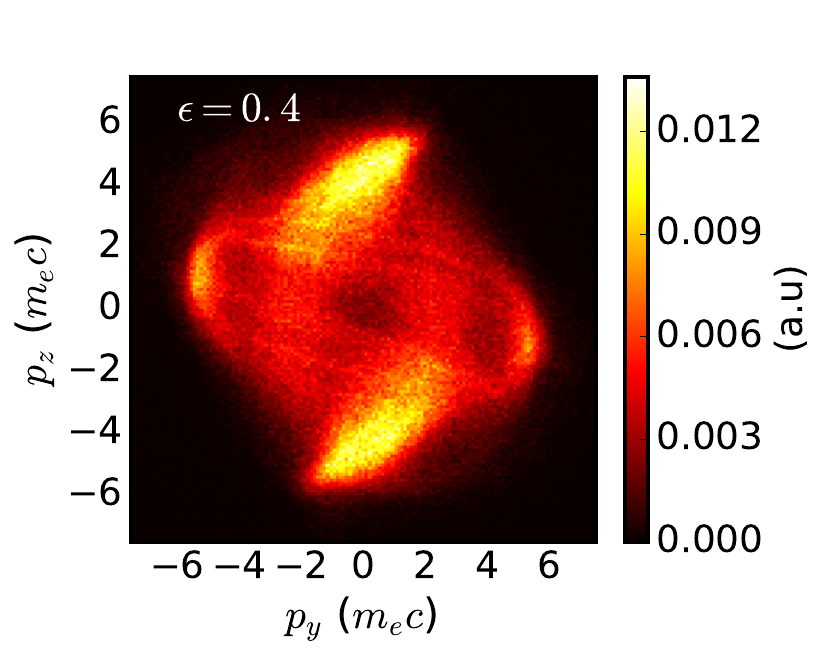}
         \label{ep3}
     \end{subfigure}
     \hfill
     \begin{subfigure}[b]{0.3\textwidth}
         \centering
         \includegraphics[width=\textwidth]{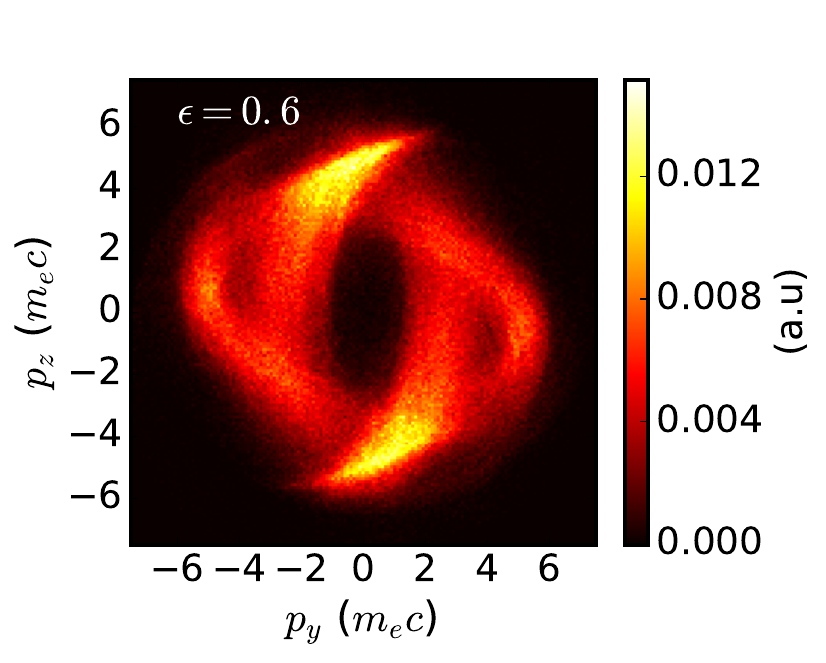}
         \label{ep7}
     \end{subfigure}
     \hfill
     \begin{subfigure}[b]{0.3\textwidth}
         \centering
         \includegraphics[width=\textwidth]{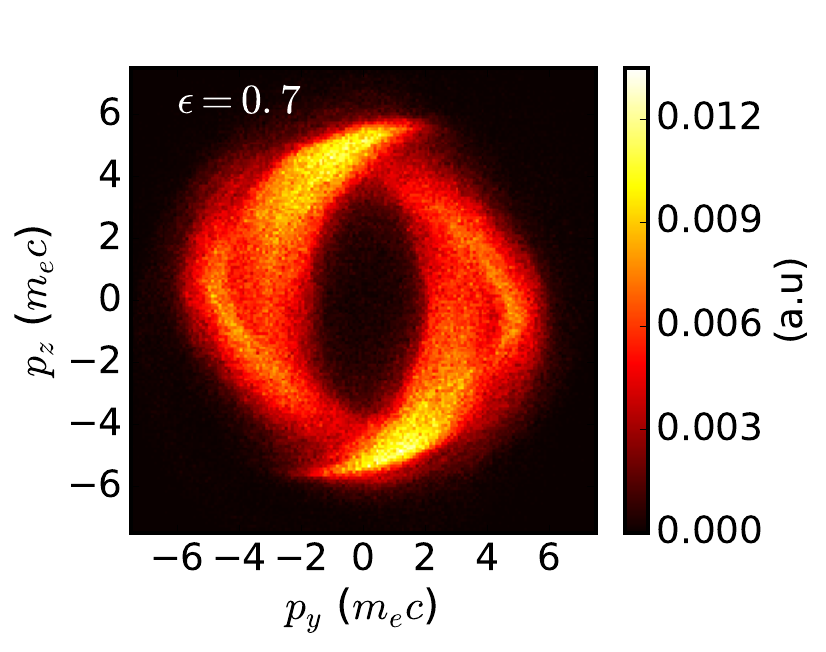}
         \label{ep8}
     \end{subfigure}
     \hfill
     \begin{subfigure}[b]{0.3\textwidth}
         \centering
         \includegraphics[width=\textwidth]{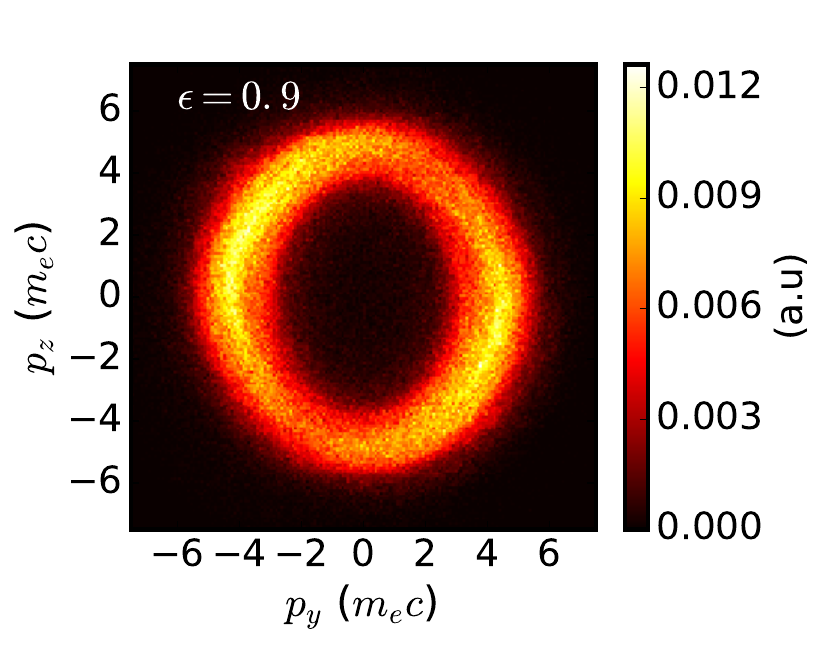}
         \label{ep9}
     \end{subfigure}
     \caption{
     2D Colormaps of the transverse momentum distribution of trapped electrons as a function of the polarization state of the laser. Compare with Fig.~\ref{figpp} for $\epsilon = 0$ and 1.}
        \label{allp}
\end{figure}
\begin{figure}
     \centering
     \begin{subfigure}[]{0.3\textwidth}
         \centering
         \includegraphics[width=\textwidth]{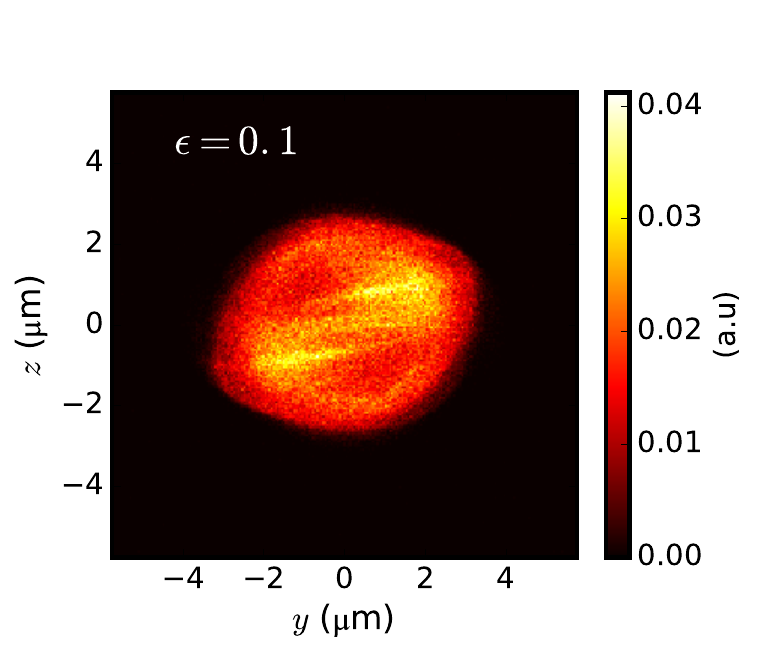}
         \label{ep1p}
     \end{subfigure}
     \hfill
     \begin{subfigure}[]{0.3\textwidth}
         \centering
         \includegraphics[width=\textwidth]{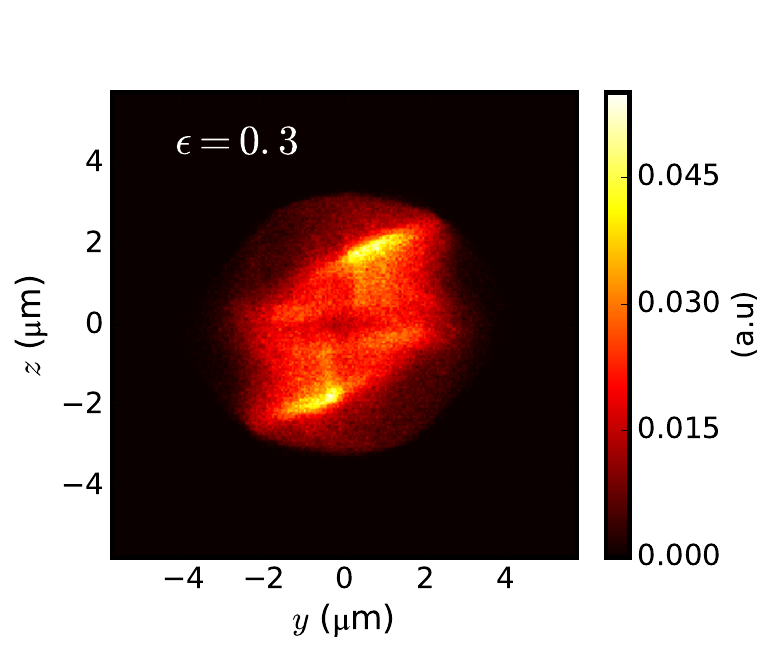}
         \label{ep2p}
     \end{subfigure}
     \hfill
     \begin{subfigure}[]{0.3\textwidth}
         \centering
         \includegraphics[width=\textwidth]{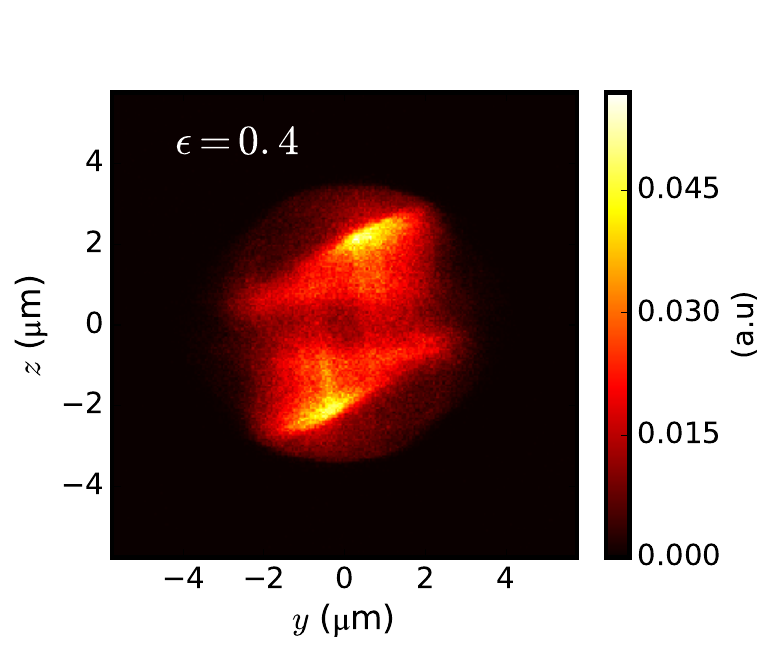}
         \label{ep3p}
     \end{subfigure}
     \hfill
     \begin{subfigure}[b]{0.3\textwidth}
         \centering
         \includegraphics[width=\textwidth]{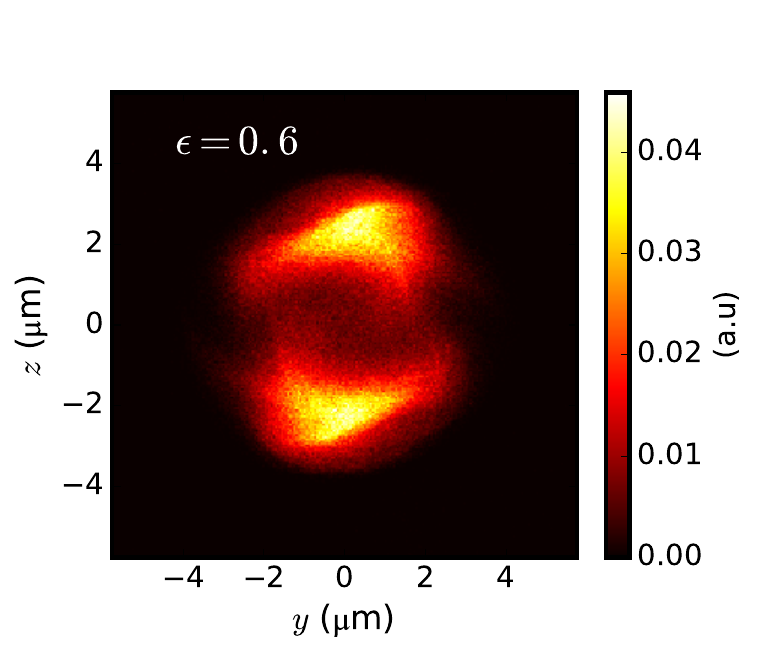}
         \label{ep7p}
     \end{subfigure}
     \hfill
     \begin{subfigure}[b]{0.3\textwidth}
         \centering
         \includegraphics[width=\textwidth]{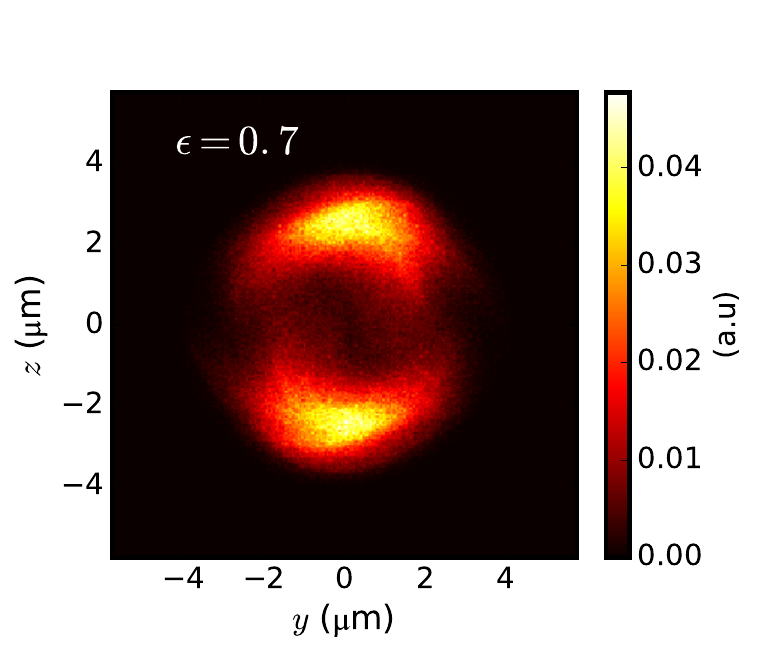}
         \label{ep8p}
     \end{subfigure}
     \hfill
     \begin{subfigure}[b]{0.3\textwidth}
         \centering
         \includegraphics[width=\textwidth]{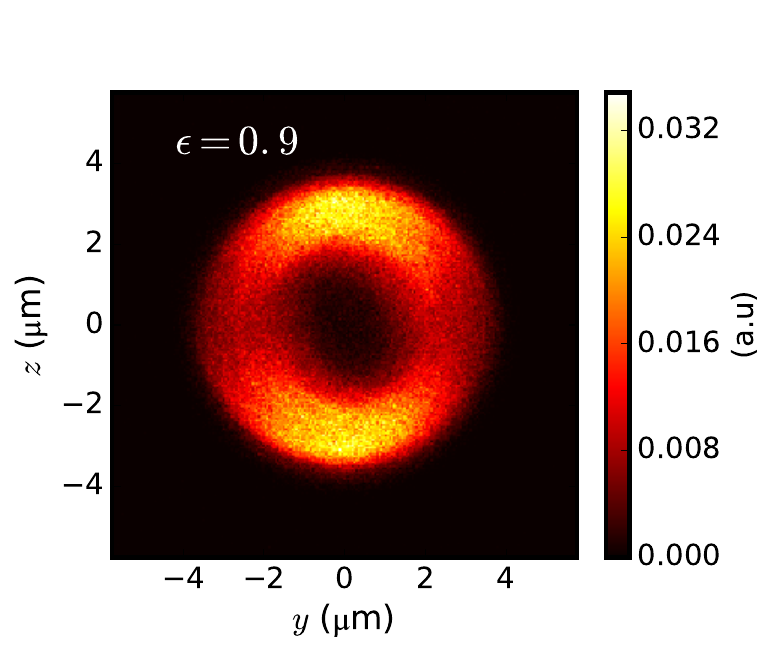}
         \label{ep9p}
     \end{subfigure}
     \caption{2D Colormaps of the transverse position distribution of trapped electrons as a function of the polarization state of the laser.}
        \label{allx}
\end{figure}

In this subsection we extend our study to the general case of elliptic polarization, i.e. we study systematically the case of intermediate values $0\leq \epsilon\leq 1$. In Figs.~\ref{allp} and \ref{allx}, 2D colormaps of the electrons transverse momentum distributions and transverse position distributions have been displayed, respectively, for different ellipticity parameters $\epsilon$. 
In both two figures we observe that for $\epsilon = 0.1$ and $0.3$, the distributions exhibits double peak situated at $z, p_z \neq 0$ and elongated along the major axis of the polarization ellipse ($y, p_y$ direction). If we increase the ellipticity further (\textit{i.e.} for $\epsilon = 0.4, 0.6, 0.7$) the momentum distributions look like two concentric ellipse oriented with an angle between them. For these values of $\epsilon$ the transverse positions distribution starts to make an ellipse with a peak oriented along the $\pm z$-direction. At $\epsilon = 0.9$, the transverse momentum and position distribution start to appear as a circular ring, but electrons are not distributed uniformly over this circular ring in contrast to the pure CP case ($\epsilon = 1$). We can qualitatively explain these behaviours from the results presented in Fig.~\ref{species} and here we consider that mostly $\ce{C}$ VI electrons are trapped inside the wake potential.
We know that the minimum value of the vector potential required for barrier suppression ionization of the $\ce{C}$ VI electron is $a_i = 1.7$. At these vector potential, the values of $\sqrt{\chi}$ and $1-\chi$ for $\ce{C}$ VI atomic electrons are 0.25 and 0.93, respectively. For $\epsilon < 0.25$, ionized $\ce{C}$ VI photo-electrons ATI momentum distribution will have a double peaked Gaussian distribution elongated along $p_y$-axis with two peaks at $p_z \neq 0$; whereas for $0.25 < \epsilon < 0.93$, the $\ce{C}$ VI electrons ATI momentum distribution will be like an ellipse.  Here at $0.6$ the elliptical behaviour is more prominent because $\epsilon = 0.6$ lies in the middle of the $\sqrt{\chi} < \epsilon < 1-\chi$ domain. 
\begin{figure}
     \centering
         \includegraphics[width=0.69\textwidth]{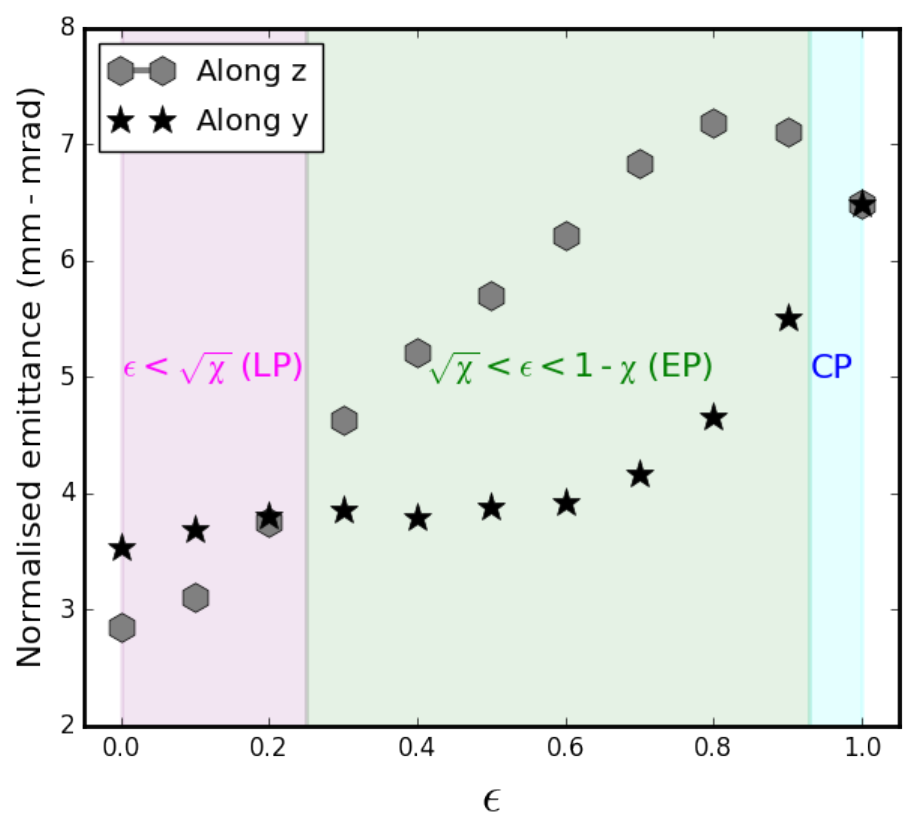}
         \caption{Dependence of the normalized transverse emittance along the $y$ and $z$ axes as function of the laser polarization $\epsilon$.}
         \label{mm}
\end{figure}

The above classification can also be seen in Fig.~\ref{mm}, where the variations of the normalized transverse emittance as a function of the laser polarization have been illustrated. Here we use the same distinction, presented in Fig.~\ref{species}, to characterize the variations of the normalized transverse emittance with respect to $\epsilon$, depicted in Fig.~\ref{mm}:

(i) LP-like regime --- in the regime $\epsilon < 0.25$ 
(shown by the magenta background color), the emittance along $z$-direction increases with $\epsilon$ as more and more photo-electrons are generated with higher $p_z$ values. But the emittance along $y$-direction doesn't increase significantly as the distribution of the ATI momentum distribution along $y$-direction after ionization in this domain doesn't change substantially \textit{w.r.t} $\epsilon$. For these values of $\epsilon$, the polarization of the laser appears as linear to the $\ce{C}$ VI electrons. (ii) EP-like regime --- in the regime $ 0.25 < \epsilon < 0.93$ (shown by the green background color), the normalized emittance along $z$-direction increases with the same trend as previous as the $z$-component of the laser field increases, while the emittance along $y$-direction starts increasing after $\epsilon = 0.6$ as the effect of ellipticity is more prominent after $\epsilon > 0.6$. For these values of $\epsilon$, the polarization of the laser appears as elliptical. Finally, (iii) CP-like regime --- for $\epsilon > 0.93$ (shown by the cyan background color, here the laser polarization can be considered as circular), where the normalized emittance along $z$ -direction starts decreasing while the emittance along $y$ direction increases until their numerical values match each other, which finally occurs at $\epsilon = 1$, \textit{i.e} for circular polarization.
\begin{figure}
         \centering
         \includegraphics[width=\textwidth]{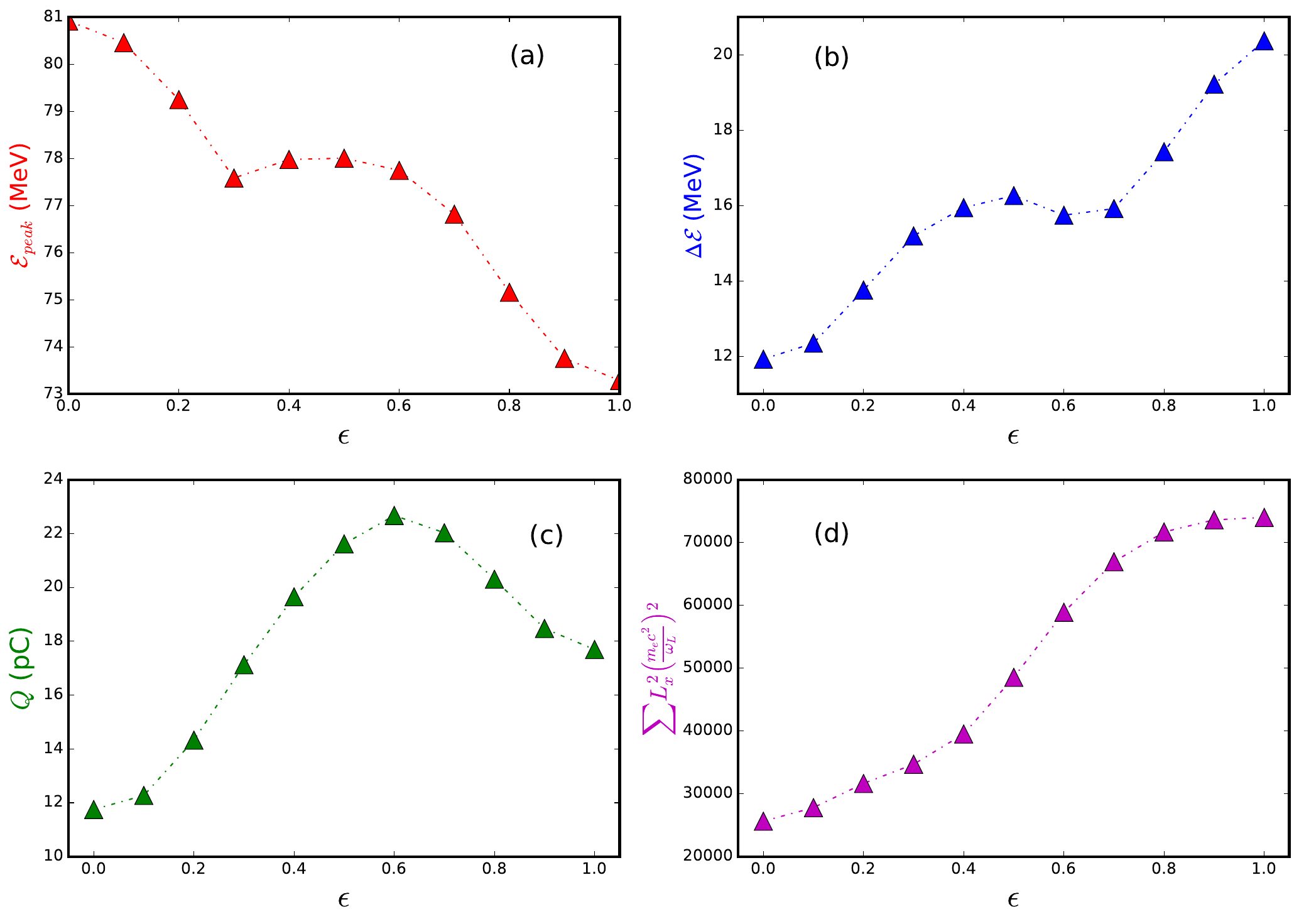}
    \caption{
    Variations of the peak energy of the beam ($\mathcal{E} _{peak}$, red points), FWHM energy spread ($\Delta \mathcal{E}$, blue points), beam charge ($Q$, green points) and the sum of the angular momentum squared ($\sum L_x^2$, magenta points) for different ellipticity parameter of the laser.}
        \label{charge}
\end{figure}

Now, in addition to the transverse emittance, the other parameters that can characterize an accelerated beam are, peak energy ($\mathcal{E} _{peak}$, where the energy distribution has its peak), FWHM energy spread ($\Delta \mathcal{E}$), beam charge ($Q$) and its angular momentum. In order to illustrate the effects of the laser polarization we have shown the variations of the above mentioned properties for different values of $\epsilon$ in Fig.~\ref{charge}. In sub-Fig.~(a), we first see that the peak energy decreases with $\epsilon$ for $0 < \epsilon < 0.3$; then for $0.3 < \epsilon < 0.6$, the peak energy remain almost constant and finally for $\epsilon$ values $> 0.6$, $\mathcal{E} _{peak}$ decreases monotonically. An opposite trend has been observed in sub-Fig.~(b), where the variation of FWHM energy spread of the accelerated electron beam has been presented as a function of the laser polarization. Here we see that $\Delta \mathcal{E}$ first increases with $\epsilon$ until $\epsilon$ reaches 0.5. Then at $\epsilon = 0.6$, the FWHM energy spread attains its minimum value and again it increases beyond $\epsilon > 0.6$. Next, the variation of beam charge $Q$ with respect to $\epsilon$ has been shown in sub-Fig.~(c). Here we can observe that, the beam charge first increases with $\epsilon$ until attains a maximum value of $\sim \unit{22}{\pico\coulomb}$ at $\epsilon = 0.6$ and then it decreases. Here, we would like mention that the variations in the peak energy ($\mathcal{E} _{peak}$), FWHM energy spread ($\Delta \mathcal{E}$) and beam charge ($Q$) doesn't change significantly (the variation occurs within the same order in the respective cases) with the laser polarization. The main important property that significantly changes with the polarization is the total angular momentum of the final accelerated beam, which has been presented in sub-Fig.~(d), by measuring the sum of the squared angular momentum projection along the beam axis ($\sum L_x^2$). In this calculation the square of the individual electrons angular momentum has been taken due to the fact that some fraction of electrons carry positive angular momentum while the rest electrons carry negative values of angular momentum. We see that $\sum L_x^2$ increases significantly with the ellipticity parameter of the laser.

\section{Betatron radiation for CP and LP cases}   \label{sect:radiation}

In this section we discuss the influence of the laser polarization on the properties of the x-ray radiation emitted by the accelerated electrons. We compute the radiation spectra emitted per unit solid angle $\Omega$ and per unit frequency $\omega$ from the bunch of the accelerated electrons, which in the far field can be written as \cite{jd} 
\begin{equation}
\frac{d^2I}{d\omega d\Omega} = \frac{e^2}{4\pi ^2  c}  \left | \int_{-\infty} ^\infty \frac{\vec{n} \times \left\lbrace  (\vec{n} - \vec{\beta}) \times \vec {\dot{\beta}} \right\rbrace }{(1 - \vec{\beta}\cdot\vec{n})^2} e^{i \omega (t - \frac{\vec{n}\cdot\vec{r}}{c})} dt\right |^2    \label{eq4}
\end{equation} 
where, $\vec{n}$ is the vector corresponding to the direction of observation and $\vec \beta = \vec v / c$ be the normalized electron velocity. The radiation is calculated by employing a post-processing code which numerically integrates Eqn.~\ref{eq4} along the trajectory of the individual electrons. Electron trajectories are obtained from PIC simulations. For radiation calculation PIC simulations have been performed up to \unit{6.07}{\pico\second} for CP and LP cases. In the post processing radiation code, radiation has been calculated along $y = 0$ (and, $z = 0$) axis using spherical polar coordinate system with polar angle $\theta_y$, ($\theta_z$), measured \textit{w.r.t} the $x$-axis (acceleration direction) in the $x$-$y$ ($x$-$z$) plane. The azimuthal angle $\phi$ is measured in the $y$-$z$ plane \textit{w.r.t} the positive $y$ axis. For a single macro-particle, the radiation along the respective axes ($y = 0$ and $z = 0$) has been calculated independently on the virtual detectors placed along the axes and the final radiation spectrum is obtained by incoherently adding all the spectra emitted by the macro-particles measured on the virtual detectors along the respective axes.
\begin{figure}
     \centering
     \includegraphics[width=0.79\textwidth]{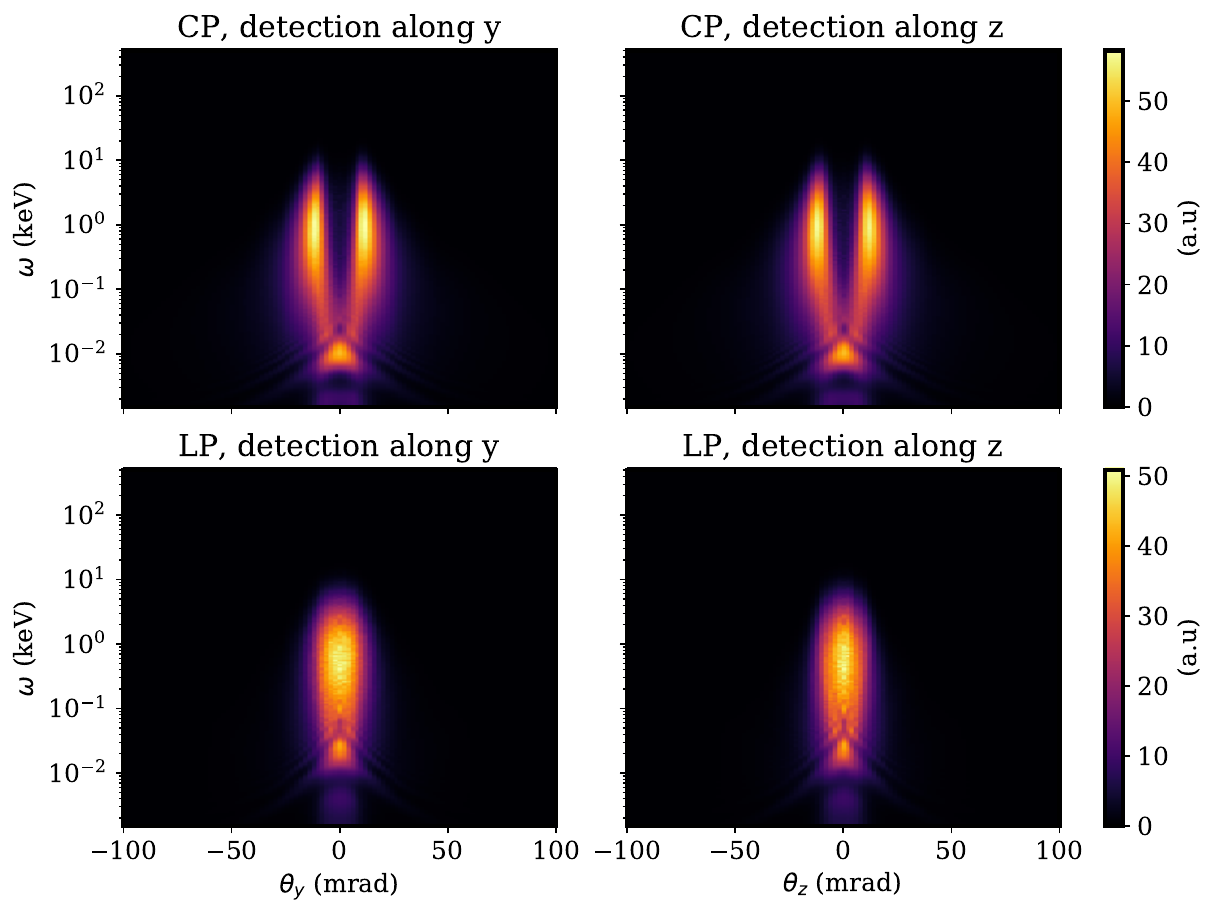}
    \caption{2D colarmaps of on axis radiation spectrum for CP (Up:Left - calculated along $y$-axis, Up:Right - calculated along $z$-axis) and LP case (Up:Left - calculated along $y$-axis, Up:Right - calculated along $z$-axis).}
    \label{irad}
\end{figure}

Fig.~\ref{irad} shows the 2D colormaps of the radiation spectra $\frac{d^2I}{d\omega d\Omega}$ along $y$ (left panel) and $z$ directions (right panel) for CP (up) and LP (down) cases. We first focus on the spatial distribution of the radiated spectrum as it can reveal the dynamics of the electron orbits along the transverse plane. We observe that for LP case the intensity distribution is maximum at $\theta _y, \theta _z \approx 0$, whereas for CP case the maximum intensity occurs at $\theta _y, \theta _z \neq 0$. These features exhibit that how the polarization state of the laser can control the spatial distribution of the emitted radiation. For LP case the higher energetic electrons present inside the beam have zero angular momentum. As a result, the orbits of the higher energetic electrons, which are contributing more in the radiation, has a two dimensional oscillating motion in the transverse plane across beam propagation axis. In this case, depending upon the ionization positions, a continuous set of initial conditions (transverse momenta and space at the instant of ionization) are possible which leads to a variety of electron trajectories with different orientation angle and betatron amplitudes. The ellipticity of the electron trajectories is also changing with time as the electrons accelerate in the forward $x$-direction. Therefore for LP case when the radiation from all the electrons incoherently added, it shows a spectrum which maximizes at $\theta _y, \theta _z \approx 0$ as the emitted radiation are directly linked to the electrons transverse motion \cite{radmain}. 
But, for CP case, electrons are injected with a well defined ATI transverse momentum which ultimately leads to a nonzero angular momentum of the mostly energetic electrons present inside the beam. Due to a nonzero angular momentum electrons move in a three dimensional helical trajectory and electron transverse trajectories are more circular as $p_y/m_ec \approx p_z/m_ec \approx p_\perp (t)/m_ec \approx a_i(\bar \gamma (t)/\gamma _i)^{1/4}$. By comparing the transverse momentum distribution (from Fig.~\ref{3pp2}) it can be mapped that, the intensity distribution will present an annular shape in the $\theta - \phi$ plane and thus the maximum radiation intensity comes from $\theta_y,\theta_z \neq 0 (\approx \unit{13}{\milli\rad})$. Thus we see that the far field radiation pattern illustrated in Fig.~\ref{irad} can be mapped from the transverse momentum distributions depicted in Fig.~\ref{fig3pp}. Similarly, from the transverse momentum distributions presented in Fig.~\ref{allp}, one can intuitively speculate the spatial distributions of the radiation spectra for other ellipticity parameters of the ionizing laser. 

\begin{table}[h]  
\caption{Radiation parameters (averaged until $\unit{6.07}{\pico\second}$)} 
\centering 
\begin{tabular}{c c c c c c c}   
\toprule
Polarization &Observation &$\gamma$   &$r_\beta$(\unit{\micro\meter})  &$\Lambda _\beta$(\unit{\micro\meter})  &$K_\beta$ 
\\ [0.5ex]  
\midrule
 LP & along $y$ &347.03 &1.03 &716.59 &3.16  \\
LP & along $z$ &347.03 &0.96 &716.59 &2.93  \\
CP & along $y$ &327.41 &1.47 &696.46 &4.36  \\
CP & along $z$ &327.41 &1.49 &696.46 &4.41  \\
\bottomrule  
\end{tabular} 
\label{t1}
\end{table}  

Now, we investigate the spectral properties of the emitted radiation. The key properties of the intensity distribution produced by a betatron source is parameterized by a fundamental dimensionless radiation parameter, called the strength parameter $K_\beta = 2\pi\gamma r_\beta/\Lambda _\beta$, where $r_\beta$ and $\Lambda _\beta$, respectively be the transverse oscillation amplitude and betatron spiral motion period. In our simulations, electrons are accelerated to a high energies with an average $\gamma$ value 327 and 347 for CP and LP cases respectively (measured until \unit{6.07}{\pico\second}). The other parameters (averaged) required to calculate the strength parameters along $y$ and $z$ directions for CP and LP cases have been presented in Table.~\ref{t1}. From this table we see that for all the cases the strength parameter is $> 1$, therefore the radiation is in the wiggler regime and synchrotron like. Now, for strength parameter larger than unity, the spectrum observed at an angle \textit{w.r.t} the $x$-axis, can be approximated by an asymptotic limit. In this limit, Eq.~\ref{eq4} can be written as \cite{jd, fa} 
\begin{equation} 
\frac{d^2I}{d\omega d\Omega}  = \frac{e^2}{3\pi^2c}\left(\frac{\omega\rho_\beta}{c}\right)^2\left(\frac{1}{\gamma ^2} + \theta ^2 \right)\left[K_{2/3}^2(\xi) + \frac{\theta ^2}{(1/\gamma ^2) + \theta ^2}K_{1/3}^2(\xi)\right]  \label{eq5}
\end{equation}
where, $K_{2/3}$ and $K_{1/3}$ are modified Bessel functions of the second kind. Here $\rho_\beta = \Lambda _\beta ^2/(4\pi^2 r_\beta)$ is the radius of curvature of the electron orbit and $\xi = \frac{\rho\omega}{3c}\left( \frac{1}{\gamma ^2} + \theta ^2\right)^{3/2}$. Now we define the critical frequency ($\omega_c$) of the emitted radiation spectra which in the practical units can be expressed as \cite{fa} 
\begin{equation} 
 \omega_c (\kilo\electronvolt) = 5.21\times 10^{-24} \gamma^2 n_e(\centi\meter^{-3})r_\beta (\micro\meter)  \label{eq6}
\end{equation}
\begin{figure}
     \centering
     \begin{subfigure}[]{0.49\textwidth}
         \centering
         \includegraphics[width=\textwidth]{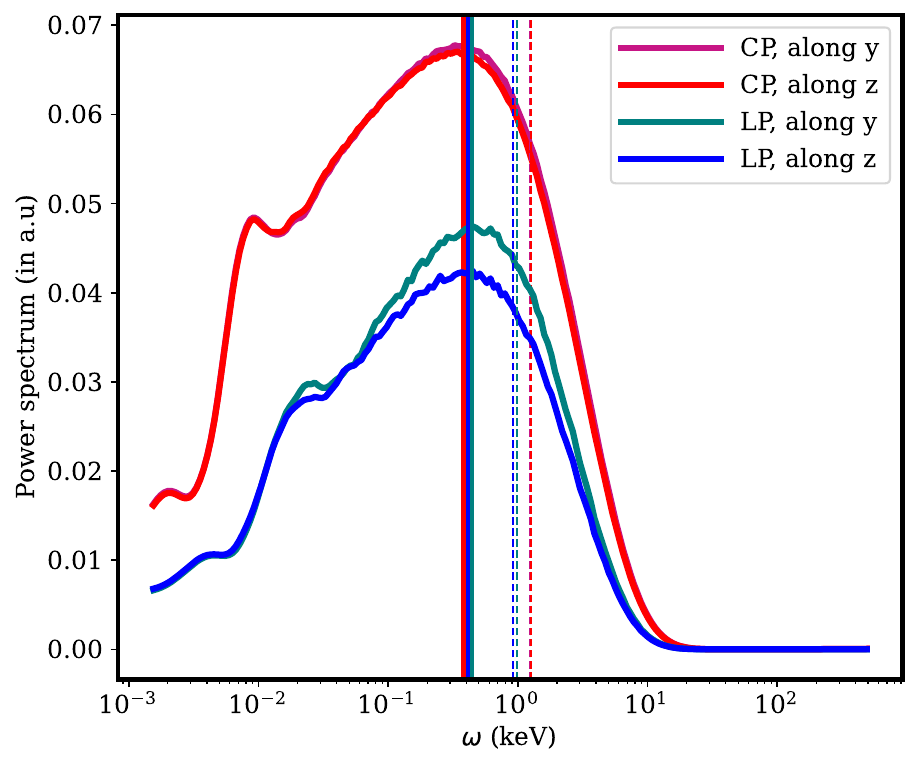}
         \caption{Variation of angle integrated radiation spectra with frequency}
         \label{angle}
     \end{subfigure}
     \hfill
     \begin{subfigure}[]{0.49\textwidth}
         \centering
         \includegraphics[width=\textwidth]{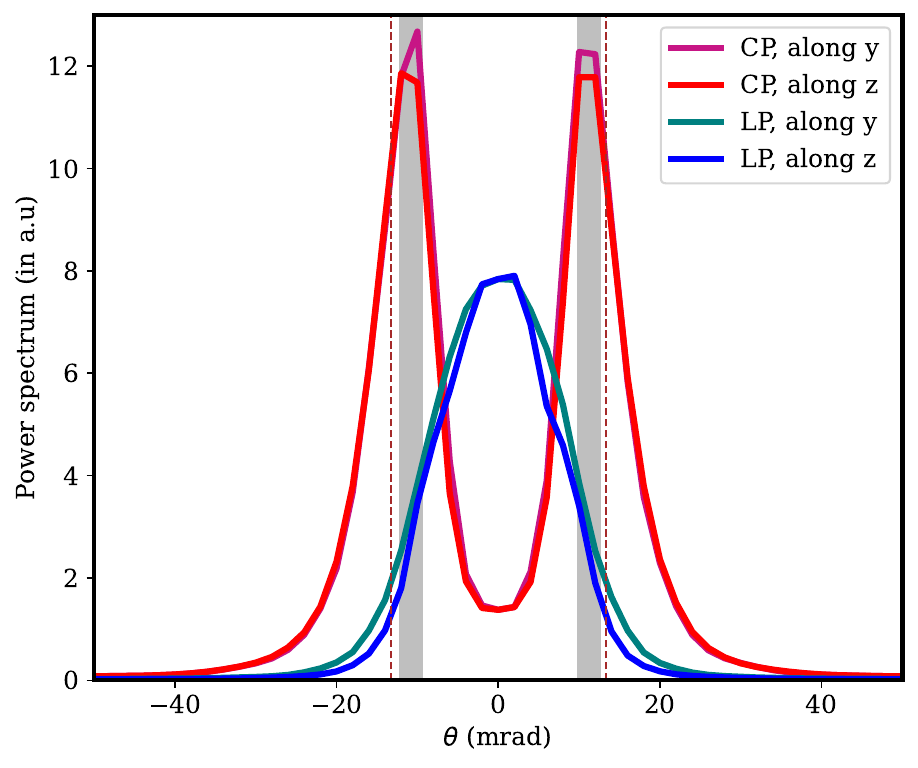}
         \caption{Variation of energy integrated radiation spectra with the angle of emission}
         \label{omega}
     \end{subfigure}
     \caption{Variation of on axis radiation spectra with frequency (left) and with the emission angle (right). Theoretically calculated critical frequencies (from Eq.~\ref{eq5}) and peak power frequencies (from Eq.~\ref{eq6}) have been shown by vertical dashed lines and solid lines on the left figure. On the right figure the vertical lines show the emission angles of the maximum intensity ($\theta = K_\beta/\gamma$, for CP case) and the angular width $\Delta \theta = 1/\gamma = \unit{3.05}{\milli\rad}$ has been presented by the grey bands.}
     \label{angle-omega}
\end{figure}
A comparison of our computational results with Eqs.~\ref{eq5} and \ref{eq6} has been shown in Fig.~\ref{angle-omega}. The left panel in this Fig.~\ref{angle-omega} shows the variation of the radiation spectra with frequency (energy) along $y$ and $z$ axis, respectively for CP and LP cases. The critical frequencies have been shown by vertical dashed lines, with the same colors of the respective cases mentioned inside the legend. For all the cases, we see that for $\omega \ll \omega_c$ spectral intensity increases with $\omega$, then it reaches maximum at a frequency near the critical frequency and finally drops exponentially to zero for $\omega > \omega_c$. From Eq.~\ref{eq5}, it is straightforward to show that the intensity distribution maximizes at $\omega_M = 0.45\omega_c$ for $\theta = 0$ (LP case) and $0.3\omega_c$ for $\theta = \unit{13}{\milli\rad}$ (CP case), respectively (as in Fig.~\ref{omega} we see that the spectrum has maximum at $\theta = 0$ for LP case and at $\theta = \unit{13}{\milli\rad}$ for CP case). The theoretically predicted $\omega_M$ values for all the cases have been illustrated by vertical solid lines. The numerical values of $\omega_c$, $\omega_M$ (and also the wavelength of the peak intensity) for all the cases have been given in Table.~\ref{t2}. From these values we find that the radiation lies in the soft x-ray regime. As the electron beam size is larger than the radiation wavelength here, the radiation is essentially incoherent.

\begin{table}[h]  
\caption{Cut-off frequency ($\omega_c$), peak power frequency ($\omega_M$) and wavelength ($\lambda _M$) at the maximum peak} 
\centering   
\begin{tabular}{c c c c c c }   
\toprule   
 Polarization &Observation &$\omega_c$ (keV)   &$\omega_M$(keV)  &$\lambda_M$  (nm) \\
\midrule
 LP & along $y$ &0.981 &0.441 &2.811   \\
LP & along $z$ &0.911 &0.410 &3.027   \\
CP & along $y$ &1.243 &0.373 &3.329   \\
CP & along $z$ &1.259 &0.377 &3.287   \\
\bottomrule   
\end{tabular} 
\label{t2}
\end{table}  
For the CP case the $\theta$ values at which the radiation is maximum can also be calculated from the parameters mentioned in Table.~\ref{t1}. From the $\gamma$ and $K_\beta$ values we find that the maximum radiation will propagate with an emission angle ($\theta = K_\beta/\gamma$) $\theta = \unit{13.3}{\milli\rad}$ along $y$-direction and $\theta = \unit{13.47}{\milli\rad}$ along $z$-direction respectively, which have been shown by vertical dashed lines in Fig.~\ref{omega}. Along both the axis the radiation will emit with an angular width $\Delta \theta = 1/\gamma = \unit{3.05}{\milli\rad}$ around the emission angle (shown by the grey patch).
\begin{figure}
     \centering
     \includegraphics[width=\textwidth]{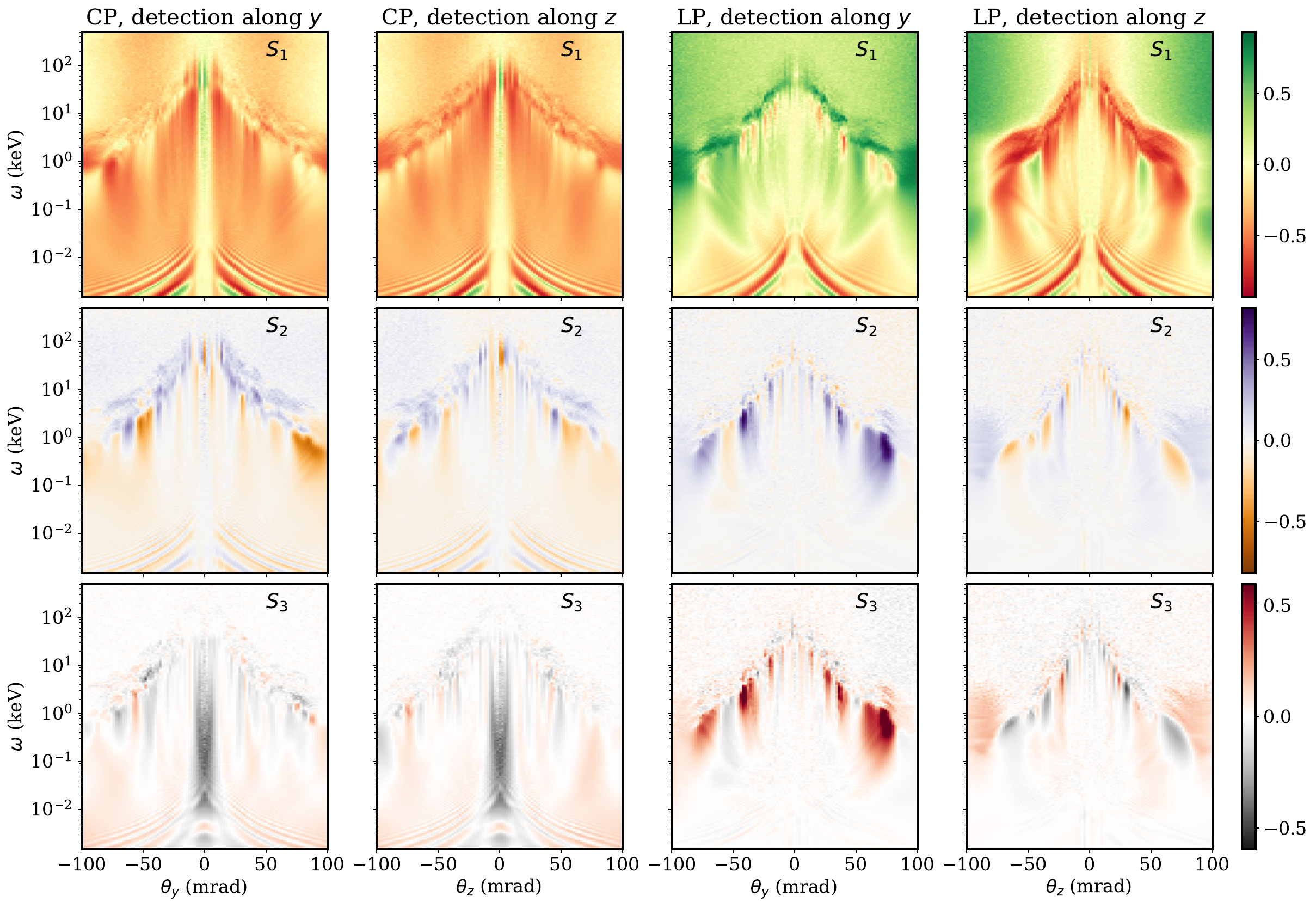}
        \caption{2D colormaps of Stokes vectors distributions of the on axis radiation spectra for CP and LP cases, calculated along $y$ and $z$ directions.}
        \label{stokes}
\end{figure}

Finally, for the complete classification of the polarization properties of the emitted radiation, we compute the Stokes parameters $\vec S = \mathrm{tr} \vec \sigma \rho / \mathrm{tr} \rho$ of the emitted photons, using the following expression \cite{BKS_book},
\begin{equation}
\rho _{ij} =  \frac{1}{2} \frac{d^2I}{d\omega d\Omega} (\mathbb{1} + \vec \sigma_{ij} \cdot \vec S)   \label{eq7}
\end{equation}
for the polarization density matrix of the emitted radiation. Here $\mathbb{1}$ is the identity matrix of order $2\times2$ and $\vec \sigma = \left\lbrace \sigma _x, \sigma _y, \sigma _z \right\rbrace$ are the Pauli matrices.
In our case the trace of the density matrix yields the intensity of emitted radiation $\mathrm{tr} \rho = d^2I/d\omega d\Omega$.
We have chosen the electric field amplitude of the emitted radiation ($\vec E_{rad}$) to characterized the polarization state and therefore we can decompose the Fourier component of $\vec E_{rad}$ over two mutually orthogonal unit vectors of circular polarization basis (say, $\vec n_1$ and $\vec n_2$), lying in a plane perpendicular to the wave propagation $\vec n$:
\begin{equation}
   \vec E_{rad} (\omega) = E_{n1}\vec n_1 + E_{n2}\vec n_2
\end{equation}
here the amplitudes of the two (orthogonal) circularly polarized electric field components are complex in general. Therefore the elements of the density matrix can be evaluated as $\rho _{ij} = E_{ni}E ^*_{nj}$. In this circular polarization basis vector representation the Stokes parameters can be expressed as: $S_1 = E_{n1}\cdot E^*_{n2} + E_{n2}\cdot E^*_{n1}$, $S_2 = -i (E_{n1}\cdot E^*_{n2} - E_{n2}\cdot E^*_{n1})$, $S_3 = E_{n1}\cdot E^*_{n1} - E_{n2}\cdot E^*_{n2}$. Here, $S_1$ represents the linear polarization along $\vec n_1$ axis or $\vec n_2$ axis; $S_2$ stands for the linear polarization at an angle $45^\degree/135\degree$ \textit{w.r.t.} $\vec n_1$ axis and $S_3$ gives a measure of circular polarization. 2D colormaps of the three Stokes parameters of the emitted radiation are presented in Fig.~\ref{stokes} for the cases of linear and circular laser polarization. The Stokes parameters have been calculated for the radiation observed by a virtual detector aligned along the $y$ and $z$-axes, respectively. We mainly focus on the maximum intensity region of the respective cases. We see that for the LP case the value of first Stokes parameter is $\approx 0.15$, along $y$-direction, and others are close to zero near the maximum intensity region ($\theta _y, \theta _z \approx 0$ and $\omega \approx \omega_M$). Contrary, for the CP case we find the radiation near the maximum intensity region ($\theta _y, \theta _z \approx \unit{13}{\milli\rad}$ and $\omega \approx \omega_M$) has nonzero values of first and third Stokes parameters. We find that the degree of linear polarization of the emitted radiation is higher than the degree of circular polarization. The degree of polarization can be measured by $S = \sqrt{(S_1^2 + S_2^2 + S_3^2)}$. We find that the maximum degree of polarization in the region of high intensity for CP case is nearly 0.8, and the radiation is polarized perpendicular to the scattering plane.
For the LP case, we find $S \approx 0.15$. By integrating over the whole spectrum, we find the betatron radiation emitted from the accelerated electrons injected via the ionization by a CP laser pulse is $36\, \%$ polarized along the scattering plane while for LP case it is nearly unpolarized, with an averaged polarization degree of $10 \, \%$. Therefore the degree of the polarization of the emitted radiation can be substantially controlled by utilizing the ionization injection method in LWFA by tuning the laser polarization.

\section{Conclusions}   \label{sect:conclusion}
In conclusion, by performing 3D PIC simulations in the nonlinear bubble regime of LWFA, we have demonstrated that the characteristics of the ionization-injected electron beams and x-ray radiation emitted by the accelerated electrons can be controlled by tuning the laser polarization. It has been found that the polarization dependent ATI momentum, gained by the ionized electrons in excess of the IP level via the above threshold ionization process, greatly affects the injected electron trajectories during the acceleration period. The effects of the laser polarization on the characteristics of the accelerated beam, such as, beam charge, peak energy, energy spread and transverse emittance have been studied in detail. It has been discovered that as the polarization of the ionizing laser gradually changes from linear to circular the higher energetic electrons, present inside the beam, carry a non-zero angular momentum with them, which can also be controlled by changing the laser polarization. By using a post-processing radiation calculation code far field radiation spectra have been calculated for LP and CP state of the laser. It has been shown that the spatial distributions and the polarization properties (Stokes vectors) of the emitted radiation for the above two cases are substantially different and thus depend on the laser polarization. The connection between the x-ray radiation and transverse momentum distribution have also been addressed for LP and CP cases. In that light the radiation properties for the other ellipticity parameters of the laser can also be predicted from the respective transverse momentum distributions \cite{radmain}, which have been presented in Fig.~\ref{allp}. Our study provides an easy and stable alternative \cite{Dpp2017} to regulate the electron beam properties and x-ray radiation in LWFAs, by employing ionization injection scheme.

\section*{Acknowledgements}
The authors gratefully acknowledge the Gauss Centre for Supercomputing e.V. (www.gauss-centre.eu) for funding this project by providing computing time through the John von Neumann Institute for Computing (NIC) on the GCS Supercomputer JUWELS at Jülich Supercomputing Centre (JSC). The research leading to the presented results received additional funding from the European Regional Development Fund and the State of Thuringia (Contract No. 2019 FGI 0013).

\section*{References}
\bibliographystyle{iopart-num}
\bibliography{CPLWFA}

\end{document}